\documentclass[oneside,11pt]{article}

\usepackage[utf8]{inputenc}
\usepackage[T1]{fontenc}

\usepackage{amsthm}
\usepackage{amsmath}
\usepackage{amsfonts}
\usepackage{amssymb}

\usepackage{graphicx} 
\usepackage{float}
\usepackage{booktabs}
\usepackage{multirow}
\usepackage{rotating}
\usepackage{array}
\usepackage{xcolor}
\usepackage{subcaption}
\usepackage[ruled]{algorithm2e}

\usepackage{breakcites}

\usepackage{enumitem}

\usepackage[top=1.5cm,bottom=2cm,right=2.5cm,left=2.5cm]{geometry}
\usepackage{titling}

\usepackage{natbib}

\usepackage{ifplatform}

\usepackage{textcomp}

\ifwindows
\fi

\ifwindows
	\usepackage{bbm}
\else
	\iflinux
		\usepackage{bbm}
	\else
		\usepackage{bbm}
	\fi
\fi

\DeclareFontFamily{OT1}{pzc}{}
\DeclareFontShape{OT1}{pzc}{m}{it}{<-> s * [1.10] pzcmi7t}{}
\DeclareMathAlphabet{\mathpzc}{OT1}{pzc}{m}{it}

\usepackage[pdftex,final,bookmarksnumbered,bookmarksopen=false,breaklinks,colorlinks]{hyperref}
\hypersetup{
	final,
	bookmarksnumbered=true,
	bookmarksopen=true,
	bookmarksopenlevel=0,
	unicode=false,          
	pdftoolbar=true,        
	pdfmenubar=true,        
	pdffitwindow=false,     
	pdftitle={Discounted optimal stopping of a Brownian bridge, with application to American options under pinning},    
	pdfdisplaydoctitle=true, 
	pdftoolbar=true,
	pdfmenubar=true,
	pdflang={English},
	pdfauthor={Bernardo D'Auria, Eduardo Garcia-Portugues, Abel Guada-Azze},     
	pdfsubject={arXiv paper},   
	pdfcreator={Abel Guada-Azze},   
	pdfproducer={Abel Guada-Azze}, 
	pdfkeywords={American option}{Brownian bridge}{Free-boundary problem}{Optimal stopping}{Option pricing}{Put-call parity}{Stock pinning}, 
	pdfnewwindow=true,      
	breaklinks=true,
	hidelinks,       
	linkcolor=black,          
	citecolor=black,        
	filecolor=magenta,      
	urlcolor=cyan           
}

\newtheorem{Remark}{Remark}
\newtheorem{Proposition}{Proposition}
\newtheorem{Lemma}{Lemma}
\newtheorem{Theorem}{Theorem}

\allowdisplaybreaks

\begin{document}

\title{Discounted optimal stopping of a Brownian bridge, with application to American options under pinning}
\setlength{\droptitle}{-1cm}
\predate{}%
\postdate{}%

\date{}

\author{Bernardo D'Auria$^{1,2}$, Eduardo Garc\'ia-Portugu\'es$^{1,2}$, and Abel Guada$^{1,3}$}

\footnotetext[1]{
	Department of Statistics, Carlos III University of Madrid (Spain).}
\footnotetext[2]{
	UC3M-Santander Big Data Institute, Carlos III University of Madrid (Spain).}
\footnotetext[3]{Corresponding author. e-mail: \href{mailto:aguada@est-econ.uc3m.es}{aguada@est-econ.uc3m.es}.}

\maketitle


\begin{abstract}
Mathematically, the execution of an American-style financial derivative is commonly reduced to solving an optimal stopping problem. Breaking the general assumption that the knowledge of the holder is restricted to the price history of the underlying asset, we allow for the disclosure of future information about the terminal price of the asset by modeling it as a Brownian bridge. This model may be used under special market conditions, in particular we focus on what in the literature is known as the ``pinning effect'', that is, when the price of the asset approaches the strike price of a highly-traded option close to its expiration date. Our main mathematical contribution is in characterizing the solution to the optimal stopping problem when the gain function includes the discount factor. We show how to numerically compute the solution and we analyze the effect of the volatility estimation on the strategy by computing the confidence curves around the optimal stopping boundary. Finally, we compare our method with the optimal exercise time based on a geometric Brownian motion by using real data exhibiting pinning. 
\end{abstract}
\begin{flushleft}
	\small\textbf{Keywords:} American option; Brownian bridge; Free-boundary problem; Optimal stopping; Stock~pinning
\end{flushleft}

\section{Introduction}

American options are a special type of vanilla option that can be considered among the most basic financial derivatives. By allowing for the possibility to exercise at any time before the day of expiration, they add a new dimension to their valuation that escapes the consolidated hedging and arbitrage-free pricing frameworks. The methodology to value an American option can be traced back to \cite{McKean1965}, which suggested transforming this problem into a free-boundary problem. However, even in the simplest case first proposed by~\cite{Samuelson1965}, where the underlying stock is modeled by a geometric Brownian motion, it took almost 40 years to reach a complete and rigorous derivation of its solution. This was given in \cite{peskir2005ontheamerican}, where it was finally proved that the free-boundary equation characterizes the optimal stopping boundary. A good historical survey on the topic can be found in~\cite{Myneni1992}. \\

The literature on the valuation of American options is considerable, and there have been many attempts to extend the classes of stochastic processes that could model the dynamics of the underlying stock. However, sometimes this has been achieved at the cost of reducing the completeness of the result. For example, \cite{Detemple2002} treats more general diffusion processes and proves that the optimal strategy satisfies the free-boundary equation, but it leaves open the proof of the uniqueness of the solution. The~more recent work \cite{Jing2012} provides a closed-form expression of the optimal stopping boundary for a fairly general class of diffusion processes by expressing it in terms of Maclaurin series. However,~\mbox{as in \cite{Detemple2002}}, it requires a boundedness assumption on the derivative of the drift term that excludes the class of Gaussian bridges. This class of processes has recently attracted attention to model situations where some future knowledge about the dynamics of the underlying assets is disclosed to the trading agent, see, e.g., \cite{PikovskyKaratzas1996}, \cite{amendinger_monetary_2003}, \cite{BiaginiOksendal2005}, and \cite{DAuriaSalmeron2020}. However, these results are mostly focused into quantifying the value of the disclosed information and do not deal with its effect on the execution strategy of a held option.\\

In the area of option pricing, a first analysis of the Brownian bridge was done in \cite{shepp1969} by exploiting a time transformation that converts the problem into one about the more tractable Brownian motion. 
Later, the applicability of this work in the problem of optimally selling a bond was highlighted in \cite{Boyce1970}. Successively, in \cite{ekstrom2009}, the authors take up the problem in \cite{shepp1969} by reframing it into the wider context of the free-boundary problem and they extend its solution to a wider class of gain functions. In particular, in \cite{ekstrom2009}, the~Brownian bridge process is presented as a possible model for financial applications under special market conditions, such as the so-called ``pinning effect''.\\

The pinning effect refers to the situation in which the price of a given stock approaches the strike price of a highly-traded option close to its expiration date. Evidence for the pinning effect was first reported in \cite{Nelken2001}, where the authors employ a bridge process to model stock pinning by tuning a geometric Brownian motion. In \cite{avellaneda2003market}, the authors postulate that the pinning behavior is mainly driven by delta-hedging of long option positions and consider, as a model for the stock price, a stochastic differential equation with a drift that pulls the price towards a neighborhood of the strike price. Later, the results in \cite{avellaneda2012mathematical} add real data evidence in support for this model. In \cite{ni2005stock}, the same assumption is validated and a comprehensive set of evidence for the pinning phenomenon is reported. The model of~\cite{avellaneda2003market} is generalized in \cite{jeannin2008modeling} by adding a diffusion term that shrinks the volatility near the strike price.\\

Motivated by these findings, we study in this paper the best strategy for executing an American put option in the presence of stock pinning. Similarly to \cite{ekstrom2009}, we model the underlying stock by a Brownian bridge. Differently, we include a discount factor in the gain function that makes the problem more realistic. This addition makes more challenging the associated optimal stopping problem, as it involves a non-perpetual option that is non-homogeneous in time. We solve this corresponding optimal stopping problem, in the spirit of \cite{peskir2005ontheamerican} and \cite{Tiziano}, by characterizing the optimal stopping boundary as the unique solution of a Volterra integral equation up to some regularity conditions.  \\

Besides contributing with the solution to this original problem, we also explore its applicability in real situations. The studied model may be too simple to be applied with real data, but it allows for computing exact solutions and to easily quantify the uncertainty about the knowledge of its parameters. For this reason, we describe an algorithm to numerically compute the optimal strategy and provide the confidence curves around the optimal stopping boundary when the stock volatility is estimated via maximum likelihood. This inferential method is potentially relevant for an investor that only has access to discrete data. \\

In addition, we test our results on a real dataset comprised by financial options on Apple and IBM equities. Our model is competitive when compared with a model based on a geometric Brownian motion and, in accordance with the motivation of our work, the best performances are obtained when the stock price exhibits a pinning-at-the-strike behavior. Finally, since our mathematical model is strongly based on the particular assumption of the pinning-at-the-strike behavior, we briefly show what effects we should expect in case of relaxing this assumption, at least from a qualitative point of view, and why a more complete analysis would require more sophisticated tools.\\

\pagebreak
We conclude by mentioning related works using similar models. In \cite{follmer_optimal_1972}, the authors tackle the non-discounted problem for a Brownian bridge with a normally-distributed ending point. More~recently, \cite{ekstrom_optimal_2020} solves the same problem for small values of the variance, finding bounds for the value function when the pinning point follows a general distribution with a finite first moment. A double stopping problem, for which the aim is to maximize the mean difference between two stopping times, is analyzed in \cite{baurdoux_chen_surya_yamazaki_2015}. The recent paper \cite{Tiziano} solves the non-discounted problem using the exponential of a Brownian bridge to model the stock prices. A Brownian bridge with unknown pinning random distribution and a Bayesian approach is advocated by \cite{GLOVER2020}. The analytical results in~\cite{ekstrom2009} are extended in \cite{DAuriaFerriero2020} by looking at a class of Gaussian bridges that share the same optimal stopping boundary. The discounted problem with a Brownian bridge and in the presence of random pinning point is addressed in \cite{Leung2018}, under regularity assumptions on the gain function that allow for an application of the standard It{\^o}'s formula (something that does not hold in our setting).\\

The rest of the paper is structured as follows. In Section \ref{sec:the_problem}, we introduce the model along with notations and definitions. Section \ref{sec:pricing} provides the theoretical results required for obtaining the free-boundary equation. Section \ref{sec:boundary_computation} deals with the problem of computing the optimal stopping boundary and quantifying the uncertainty associated with the estimation of the stock volatility. Section~\ref{sec:real_data} compares our method with the optimal exercise time based on a geometric Brownian motion by using real data exhibiting various degrees of pinning. Section \ref{sec:pinning-at-any-point} comments about the relaxation of the pinning-at-the-strike assumption and, finally, Section~\ref{sec:concluding_remarks} offers some conclusions.\\

Proofs and technical lemmas required to back up Section \ref{sec:pricing} are relegated to Appendix \ref{sec:sup_A} and \ref{sec:sup_B} respectively. Supplementary Materials are available online (see Section \ref{supplement}) for the implementation and reproducibility of the methods and simulations exposed in Section \ref{sec:boundary_computation}.

\section{Problem setting}
\label{sec:the_problem}

We introduce next the model of the financial asset and we define the optimal stopping problem whose solution constitutes the best strategy to exercise an American put option based on that asset.\\

We assume that the financial option has a strike price $S > 0$ and a maturity date $T>0$. To~model the pinning effect, we use a Brownian bridge process for the dynamics of the underlying asset. Indeed,~this process may be seen as a Brownian motion conditioned to terminate at a known terminal value, that in our case is fixed to the strike price $S$ (see Section~\ref{sec:pinning-at-any-point} for a discussion on the relaxation of this assumption). That is, by calling $X^{[t, T]}=(X_{t+s}, 0 \leq s \leq T - t)$ to the asset price process, with~$0 \leq t<T$, we assume it satisfies the SDE:
\begin{align}
	\label{eq:BB_SDE}
	X_t &= x , \quad\quad \mathrm{d}X_{t+s} = \displaystyle{\frac{S-X_{t+s}}{T-t-s}}\,\mathrm{d}s + \sigma\,\mathrm{d}W_{s}, 
\end{align}
with $0 \leq s \leq T - t$ or, equivalently, it has the explicit expression
\begin{align}
	\label{eq:BB_explicit}
	X_{t+s} &= x \, \frac{T-t-s}{T-t} + S\frac{s}{T-t} + \sigma\sqrt{\frac{T-t-s}{T - t}}W_s, 
\end{align}
again with $0 \leq s \leq T - t$ and where, in both equations, $(W_s, 0 \leq s \leq T - t)$ denotes a standard Brownian motion. To emphasize that the process almost surely satisfies the relation $X_t = x$, we will use the notation $\mathbb{P}_{t, x}$ and $\mathbb{E}_{t, x}$ to denote the corresponding probability and mean operators.\\

Denoting by $G(x) = (S - x)^{+}$ the \emph{gain function} of the put option and by $\lambda \geq 0$ the discounting rate, we can finally write the optimal expected reward for exercising the American option as the \emph{Optimal Stopping Problem} (OSP)
\begin{align}\label{eq:OSP_AmPut_BB}
	V(t, x) = \sup_{0\leq\tau\leq\ T-t}\mathbb{E}_{t, x}\left[e^{-\lambda\tau}G(X_{t + \tau})\right] .
\end{align}

The function $V$ is called the \emph{value function} and the supreme above is taken over all the stopping times $\tau$ of $X^{[t,T]}$ with respect to its natural filtration $(\mathcal{F}_s)_{s=0}^T$.\\

Under mild conditions, namely $V$ being lower semi-continuous and $G$ upper semi-continuous (see \citet[Corollary 2.9]{goran-optimal}), it is guaranteed that the supremum in \eqref{eq:OSP_AmPut_BB} is achieved. The \emph{Optimal Stopping Time} (OST), $\tau^*(t, x)$, is defined as the smallest stopping time attaining the supremum \eqref{eq:OSP_AmPut_BB} and can be characterized as the hitting time of a closed set~$D$, referred to as the \textit{stopping set}. Since these conditions on $V$ and $G$ are satisfied in our settings (see \citet[Remark 2.10]{goran-optimal}), we can write 
\begin{align}\label{eq:OST}
	\tau^*(t,x):=\inf\{0\leq s\leq T-t : X_{t+s}\in D \mid X_{t}=x \},
\end{align}
where $D$ is defined as 
\begin{align}\label{eq:stopping_set}
	D := \left\{(t, x)\in [0,T]\times\mathbb{R} : V(t, x) = G(x)\right\}.
\end{align}

We then define the \textit{continuation set} $C$ as the complement of the set $D$ and we denote by $\partial C$ its~boundary.\\

The OST, defined in \eqref{eq:OST}, can be interpreted as the best exercise strategy for the American option,
and it allows for rewriting the value function $V$ in the simplified form 
\begin{align}\label{eq:Val_fun}
	V(t, x) = \mathbb{E}_{t, x}\left[e^{-\lambda \tau^*(t, x)}G\left(X_{t + \tau^*(t, x)}\right)\right].
\end{align}

To solve the OSP given in \eqref{eq:OSP_AmPut_BB},
we follow the well-known approach of reformulating it as a free-boundary problem, for the unknowns $V$ and $\partial C$. The latter is commonly called the \emph{Optimal Stopping Boundary}~(OSB). Our OSP is a finite-horizon problem that involves a time non-homogeneous process, with its associated free-boundary problem being 
\begin{subequations} \label{eq:free-boundary}
	\begin{align}
		\partial_tV + \mathbb{L}_{X}V &= \lambda V && \text{on } C , \label{eq:free-boundary1}\\
		V &> G && \text{on } C , \label{eq:free-boundary2}\\
		V &= G && \text{on } D , \label{eq:free-boundary3}\\
		\partial_xV &= \partial_xG && \text{on } \partial C , \label{eq:free-boundary4}
	\end{align}
\end{subequations}
where $\mathbb{L}_X$ is the infinitesimal generator of the Brownian bridge $X^{[0, T]}$. Given a suitably smooth function $f:[0, T]\times\mathbb{R} \rightarrow \mathbb{R}$, the application of the operator $\mathbb{L}_X$ to it returns the function
\begin{align}\label{eq:inf_gen}
	\left(\mathbb{L}_{X}f\right)(t, x) = \frac{S - x}{T - t}\partial_xf(t, x) + \frac{\sigma^2}{2}\partial_{x^2}f(t, x) . 
\end{align}

Equations \eqref{eq:free-boundary1}, \eqref{eq:free-boundary2}, and \eqref{eq:free-boundary3} easily come from the definitions of $D$, $C$, and $\tau^*(t, x)$ (see~Proposition~\ref{pr:V} below), whereas \eqref{eq:free-boundary4}, generally known as the \textit{smooth fit condition}, depends on how well-behaved the OSB is for the underlying process. The regularity of the OSB is an important factor in finding and characterizing the solution of the problem itself, and for this reason we will study it in detail in later sections. An in-depth survey on the optimal stopping theory that exploits the free-boundary approach can be found in \cite{goran-optimal}.

\section{Optimally exercising American put options for a Brownian bridge}\label{sec:pricing}

We present in this section the main result, consisting of the solution of the problem \eqref{eq:OSP_AmPut_BB}. In~particular, we solve the free boundary problem defined in \eqref{eq:free-boundary} by showing that the OSB can be written in terms of a function $b(t)$ that is $\partial C = \left\{(t, b(t)): t \in [0,T] \right\}$, and that this function can be computed as the solution to a Volterra integral equation.\\

From an applicative perspective, the function $b$ defines the optimal strategy to follow in order to maximize the profit from the execution of the American put option. It is the best to exercise the option the first time the price of the underlying financial asset crosses at time $t\in[0,T]$ the level~$b(t)$.

\begin{Theorem}\label{thm:OSP_sol}
	The optimal stopping time in \eqref{eq:OST} can be written as
	$$
	\tau(t, x) = \inf \{s\in [0, T - t] : X_{t + s} \leq b(t + s)\} , \quad 
	x\in\mathbb{R} , \,\,\,
	0 \leq t \leq T ,
	$$   
	where the function $b$ is defined as the unique solution, among the class of continuous functions of bounded variation lying below $S$, 
	of the integral equation
	\begin{align}\label{eq:b_volterra}
		b(t) = S - \int_t^{T} K_{\sigma, \lambda}(t, b(t), u, b(u))\,\mathrm{d}u .
	\end{align}
	
	In addition, the value function $V$ in \eqref{eq:Val_fun} can be expressed as
	\begin{align}\label{eq:V_volterra}
		V(t,x) = \int_t^{T}K_{\sigma, \lambda}(t, x, u, b(u))\,\mathrm{d}u.
	\end{align}
	
	The kernel $K_{\sigma, \lambda}$ in \eqref{eq:b_volterra} and \eqref{eq:V_volterra} is defined as
	\begin{equation}\label{eq:K}
		\begin{aligned}
			K_{\sigma, \lambda}(t, x_1, u, x_2) := e^{-\lambda (u - t)} \frac{1+\lambda(T-u)}{T - u} 
			& \Big[(S - \mu(t, x_1, u))\Phi(z_{\sigma}(t, x_1, u, x_2)) \\
			& \quad\quad+ \nu_{\sigma}(t, u)\phi(z_{\sigma}(t, x_1, u, x_2)) \Big],
		\end{aligned}
	\end{equation}
	where $\Phi$ and $\phi$ are, respectively, the distribution and the density functions of a standard normal random variable,
	\begin{align}
		\mu(t, x, u) &= x\frac{T - u}{T - t} + S\frac{u - t}{T - t}, \label{eq:BB_mean}\\ 
		\nu_{\sigma}(t, u)	&= \sigma\sqrt{\frac{(u - t) (T - u)}{T - t}}, \label{eq:BB_var} 
	\end{align}
	and $z_{\sigma}(t, x_1, u, x_2)	= (x_2 - \mu(t, x_1, u))/\nu_{\sigma}(t, u)$. 
\end{Theorem}

The proof of the theorem makes use of some important partial results that we state in the following propositions. All the proofs are deferred to Appendix \ref{sec:sup_A}.\\

The next result sheds some light on the shapes of the sets $D$ and $C$, by showing that their common border can be expressed by means of a function $b$ that satisfies some regularity conditions. In the proof, we focus on regions where it is easy to prove that the value function either exceeds or equals the immediate reward, thus revealing subsets of $C$ and $D$, respectively. Those regions come from the fact that $G$ is null above $S$ and positive below, the paths of the process decrease with $x$ (for a fixed realization), and $V$ is non-increasing with respect to $x$, $t$, and $\lambda$. 

\begin{Proposition}\label{pr:b_existence}
	There exists a non-decreasing right-continuous function $b:[0,T]\rightarrow \mathbb{R}$ such that $b(t) < S$ for all $t\in[0,T)$, $b(T) = S$, and $D = \left\{(t, x)\in[0,T]\times\mathbb{R}: x \leq b(t)\right\}$.
\end{Proposition}

The next proposition analyzes the regularity properties of the value function and proves that the smooth fit condition holds. It uses an extended version of the Itô's formula (see Lemma \ref{lm:change_var2}) to prove the monotonicity property that exploits the regularity properties of the function $b$ proved in Proposition~\ref{pr:b_existence}. We later use these results in Proposition \ref{pr:b_continuity} to show the continuity of $b$. Part \ref{pr:V.i} comes from standard arguments on parabolic partial differential equations in conjunction with the Markovian property of the Brownian bridge. The rest of the proposition employs different methods, but they all rely on the fact that the OST for a pair $(t, x)$ is sub-optimal under different initial~conditions. 

\begin{Proposition}\label{pr:V}
	The value function $V$ defined in \eqref{eq:OSP_AmPut_BB} satisfies the following conditions:
	\begin{enumerate}[label=(\roman{*}), ref=(\textit{\roman{*}}),leftmargin=2.3em,labelsep=4mm]
		\item $V$ is $\mathcal{C}^{1,2}$ on $C$ and on $D$, and $\partial_tV + \mathbb{L}_{X}V = \lambda V$ on $C$. \label{pr:V.i}
		\item $x\mapsto V(t,x)$ is convex and strictly decreasing for all $t\in[0,T]$. Moreover, \label{pr:V.ii}
		\begin{align}\label{eq:V_x}
			\partial_xV(t,x) = -\mathbb{E}\left[e^{-\lambda\tau^*(t, x)}\frac{T-t-\tau^*(t,x)}{T-t}\right].
		\end{align} 
		\item The smooth fit condition holds, i.e., $\partial_xV(t,b(t)) = -1$ for all $t\in[0,T]$. \label{eq:smooth.fit}
		\item $t\mapsto V(t,x)$ is non-increasing for all $x\in\mathbb{R}$. \label{pr:V.iv}
		\item $V$ is continuous. \label{pr:V.v} 
	\end{enumerate}
\end{Proposition}

From the previous result, we are able to get a stronger result on the regularity of $b$ 
that is used to characterize $b$ as the unique solution of the Volterra equation. We prove it by assuming that the OSB allows discontinuities of the first type and reaching a contradiction. Then, since $b$ is non-increasing and finite, it does not allow discontinuities of the second type either and must be continuous. 

\begin{Proposition}\label{pr:b_continuity}
	The optimal stopping boundary $b$ for the problem \eqref{eq:OSP_AmPut_BB} is continuous.
\end{Proposition}

Finally, the next proposition shows that the OSB satisfies the Volterra integral Equation \eqref{eq:b_volterra}
and that it is the only solution up to some regularity conditions. The proof follows well-known procedures based on probabilistic arguments (see \cite{peskir2005ontheamerican}) rather than relying on integral equation's theory, which~usually uses some variation of the contraction mapping principle.

\begin{Proposition}\label{pr:uniqueness}
	The optimal stopping boundary $b$ for the problem \eqref{eq:OSP_AmPut_BB} can be characterized as the \emph{unique} solution of the type two nonlinear Volterra integral Equation \eqref{eq:b_volterra}, within the class of continuous functions of bounded variation  $c:[0,T]\rightarrow\mathbb{R}$ such that $c(t) < S$ for all $t \in (0, T)$.
\end{Proposition}

\begin{Remark}\label{rem:put-call}
	All the results in this section have their own analog when it comes to optimally exercising American \emph{call} options, that is, when the gain function in \eqref{eq:OSP_AmPut_BB} is substituted with $G(x) = (x - S)^+$. Indeed, exploiting the symmetry of the Brownian bridge and the gain functions, it is easy to check that the relation $b_c(t) = 2S - b_p(t)$ holds, where $b_c$ and $b_p$ stand for the OSBs for the call and the put option, respectively.
\end{Remark}

\begin{Remark}\label{rem:lambda0}
	Setting $\lambda = 0$, we can recover the OSB that maximizes the mean of a Brownian bridge, and using the results in \cite{shepp1969}, \cite{ekstrom2009}, and \cite{ernst_revisiting_2016}, we get an explicit expression for the OSB that is $b_0(t) = S - \sigma B\sqrt{T - t}$, with  $B \approx 0.8399$. 
\end{Remark}

\section{Boundary computation and inference}
\label{sec:boundary_computation}

\subsection{Solving the free-boundary equation}\label{subsec:solving_the_fbe}

The lack of an explicit solution for \eqref{eq:b_volterra} requires a numerical approach to compute the OSB. Let~$(t_{i})_{i = 0}^{N}$ be a grid in the interval $[0, T]$ for some $N \in \mathbb{N}$. The method we consider builds on a proposal by \cite{pedersen2002nonlinear}. They suggested to approximate the integral in \eqref{eq:b_volterra} by a right Riemann sum, hence~enabling the computation of the value of $b(t_i)$, for $i = 0, \dots, N - 1$, by using only the values $b(t_j)$, with~$j = i + 1, \dots, N$. Therefore, by knowing the value of the boundary at the last point $(b(t_{N}) = b(T) = S)$, one can obtain its value at the second last point $b(t_{N - 1})$ and recursively construct the whole OSB evaluated at $(t_{i})_{i = 0}^{N}$. \\

Under our settings, the right Riemann sum is no longer a valid option because we know from~\eqref{eq:K} that, depending on the shape of the boundary $b$ near the expiration date, $K(t_{i}, b(t_{i}), u, b(u))$ could explode as $u\rightarrow T$, so we cannot evaluate the kernel $K$ at the right point in the last subinterval $(t_{N-1}, T]$. To deal with this issue, we employ a right Riemann sum approximation along all the subintervals except the last one, ending up with the following discrete version of the Volterra integral Equation \eqref{eq:b_volterra}:
\begin{align}\label{eq:riemann}
	b(t_{i}) \approx S - \sum_{j = i + 1}^{N - 1}(t_{j} - t_{j})K_{\sigma, \lambda}(t_{i}, b(t_{i}), t_{j}, b(t_{j})) - I(t_{i}, t_{N - 1}),
\end{align}
for $i = 0,1,\dots, N - 1$, where $I(t_{i}, t_{N - 1}) := \int_{t_{N - 1}}^{T}K_{\sigma}(t_{i}, b(t_{i}), u, b(u))\,\mathrm{d}u$. It can be shown that $ 0\leq I(t_{i}, t_{N - 1})\leq H(t_{i}, t_{N - 1})$, where
\begin{align*}
	H(t_{i}, t_{N - 1}) :=&\; e^{-\lambda (t_{N - 1} - t_i)}\int_{t_{N - 1}}^{T}(1 + \lambda(T - u))\left(\frac{S - b(t_i)}{T - t_i} + \sigma\sqrt{\frac{1}{2\pi(T - u)}}\right)\,\mathrm{d}u \\
	=&\; e^{- \lambda (t_{N-1} -  t_i)} \left((S - b(t_i))\frac{T - t_{N - 1}}{T - t_i}\left(1 + \frac{\lambda}{2}(T - t_{N - 1})\right) + \right. \\
	& \hspace{3cm} \left.  \sigma\sqrt{\frac{2(T - t_{N - 1})}{\pi}}\left(1 + \frac{\lambda}{3}(T - t_{N - 1})\right)\right),
\end{align*}
by using \eqref{eq:BB_mean} and \eqref{eq:BB_var}, the form of the kernel \eqref{eq:K}, and the fact that $\Phi(x) \leq 1$ and $\phi(x) \leq (2\pi)^{-1/2}$ for all $x\in\mathbb{R}$. Therefore, $I(t_{i}, t_{N - 1})\approx H(t_{i}, t_{N - 1})/2$ can be seen as a reasonable approximation, admitting an upper bound for the error $\varepsilon(t_i, t_{N - 1}) := \vert H(t_{i}, t_{N - 1})/2 - I(t_{i}, t_{N - 1}) \vert$, namely $\varepsilon(t_i, t_{N - 1})\leq H(t_{i}, t_{N - 1})/2$. Moreover, $H(t_{i}, t_{N - 1}) = \mathcal{O}(\sqrt{T - t_{N - 1}})$ as $t_{N - 1}\rightarrow T$. After substituting $I(t_i, t_{N - 1})$ for $H(t_{i}, t_{N - 1})/2$ in \eqref{eq:riemann}, we get
\begin{align}
	b(t_{N - 1}) &\approx \left(\frac{1}{2} - \frac{\lambda}{4}(T - t_{N - 1})\right)^{-1} \nonumber \\
	&\times\left(\frac{S}{2}\left(1 - \frac{\lambda}{2}(T - t_{N - 1})\right) - \sigma\sqrt{\frac{T - t_{N - 1}}{2\pi}}\left(1 + \frac{\lambda}{3}(T - t_{N - 1})\right)\right), \label{eq:second-last_t_approx_1}  \\
	b(t_i) &\approx \left(1 -  \frac{1}{2}e^{-\lambda(t_{N - 1} - t_i)}\left(1 + \frac{\lambda}{2}(T - t_{N - 1})\right)\frac{T - t_{N - 1}}{T - t_i}\right)^{-1} \nonumber\\
	&\times \left(S - \sum_{j = i + 1}^{N - 1}(t_{j} - t_{j})K_{\sigma, \lambda}(t_{i}, b(t_{i}), t_{j}, b(t_{j})) \right. \nonumber \\
	& \hspace{0.8cm} -\frac{1}{2}e^{-\lambda(t_{N-1} - t_{i})}\left(S\frac{T -t_{N - 1}}{T - t_i}\left(1 + \frac{\lambda}{2}(T - t_{N - 1})\right) \right. \nonumber \\  
	& \left.\left. \hspace{3.8cm} +\ \sigma\sqrt{\frac{2(T - t_{N - 1})}{\pi}}\left(1 + \frac{\lambda}{3}(T - t_{N - 1})\right)\right)\right). \label{eq:second-last_t_approx_2} 
\end{align}

The procedure for computing the estimated boundary according to the previous approximations is laid down in Algorithm \ref{alg:OSB}. From now on, we will use $\tilde{b}$ to denote the cubic-spline interpolating curve that goes through the numerical approximation of the boundary at the points $(t_{i})_{i=0}^{N}$ via Algorithm \ref{alg:OSB}.\\

\begin{algorithm}[H]
	\textbf{Input: } $S$, $\lambda$, $(t_{i})_{i = 0}^{N}$, $\delta$ \\
	\textbf{Output: } $(\tilde{b}(t_{i}))_{i = 0}^{N}$ \\
	\textbf{Code: }\\
	$\tilde{b}(T) \leftarrow S$\\
	Update $\tilde{b}(t_{N - 1})$ according to \eqref{eq:second-last_t_approx_1} \\
	\For{$i = N - 2$ \KwTo $0$}{
		$\tilde{b}(t_i) \leftarrow \tilde{b}(t_{i + 1})$ \\
		$\varepsilon \leftarrow 1$\\
		\While{$\varepsilon > \delta$}{
			$\tilde{b}_{\mathrm{old}}(t_{i}) \leftarrow \tilde{b}(t_{i})$	\\
			Update $\tilde{b}(t_{i})$ according to \eqref{eq:second-last_t_approx_2} \\
			$\varepsilon \leftarrow |\tilde{b}_{\mathrm{old}}(t_{i}) - \tilde{b}(t_{i})|/|\tilde{b}_{\mathrm{old}}(t_{i})|$ \\
		}
	}
	\caption{\small Optimal stopping boundary computation}
	\label{alg:OSB}
\end{algorithm}
\vspace{12pt}

Recall from Remark \ref{rem:lambda0} that our OSB for $\lambda=0$ takes the form $b_0(t) = S - B\sigma\sqrt{T - t}$. 
Having the explicit form of $b_0$ allows us to validate the accuracy of Algorithm \ref{alg:OSB} and to tune up its parameters. We~empirically determined that $\delta= 10^{-3}$ offers a good trade-off between accuracy and computational time. This value was considered every time Algorithm \ref{alg:OSB} was employed. \\

\begin{figure}[h!]
	\centering
	\begin{subfigure}[b]{0.45\textwidth}
		\includegraphics[width = \textwidth]{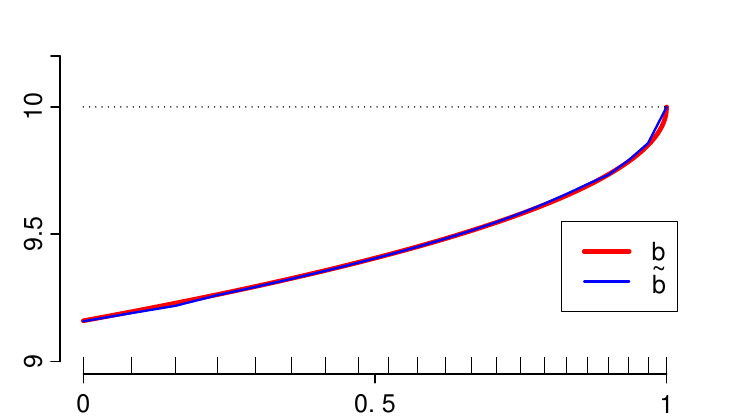}
		\subcaption{$N = 20$}
	\end{subfigure}
	\begin{subfigure}[b]{0.45\textwidth}
		\includegraphics[width = \textwidth]{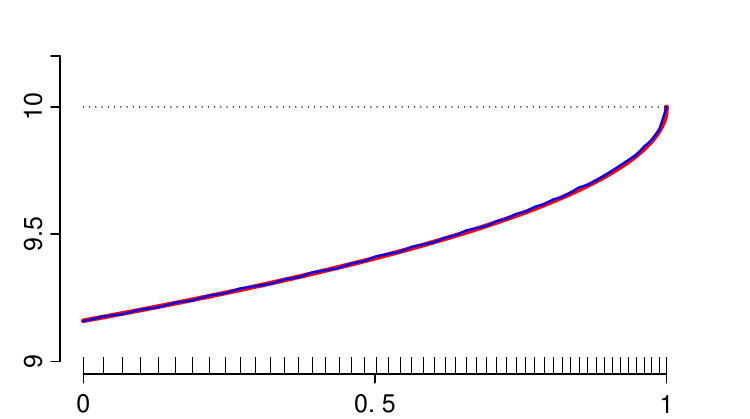}
		\subcaption{$N = 50$}
	\end{subfigure}
	\begin{subfigure}[b]{0.45\textwidth}
		\includegraphics[width = \textwidth]{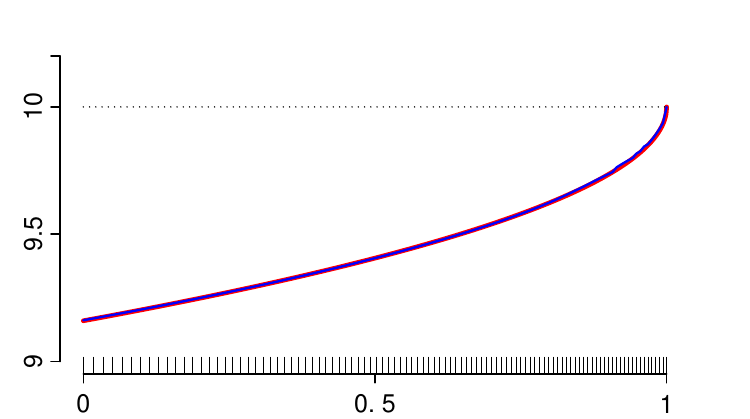}
		\subcaption[c]{$N = 100$}
	\end{subfigure}
	\begin{subfigure}[b]{0.45\textwidth}
		\includegraphics[width = \textwidth]{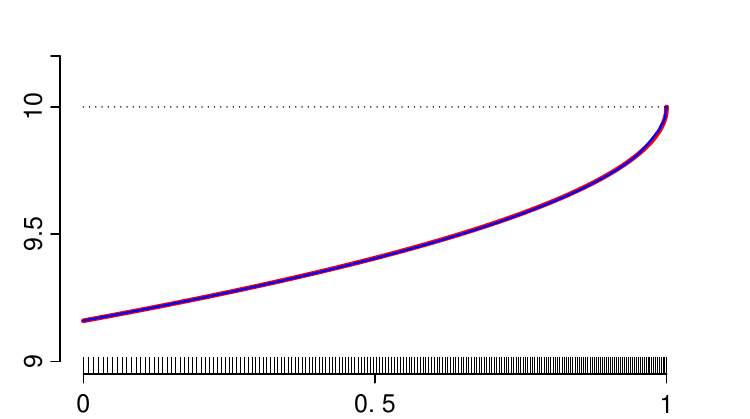}
		\subcaption[d]{$N = 200$}
	\end{subfigure}
	\vspace{6PT}
	
	\caption{\small Boundary estimation via Algorithm \ref{alg:OSB} for different partition sizes and for the parameters $S = 10$, $T = 1$, $\lambda = 0$, and $\sigma = 1$. The estimation becomes more accurate as the partition gets thinner.}
	\label{fig:b_est}
\end{figure}

We decided to use a logarithmically-spaced grid that is $t_i = \log(1 + \frac{i}{N}(e^{T} - 1) ),\ i = 0, \dots, N$, with $N = 200$, after systematically observing that uniform partitions tend to misbehave near the expiration date $T$. In addition, it is preferable that the partition gets thinner close to $T$ in a smooth way. Figure \ref{fig:b_est} shows how precise the Algorithm \ref{alg:OSB} is by comparing the computed boundary $\tilde{b}_0$ versus its explicit form for $S = 10$, $T = 1$, $\lambda = 0$, and $\sigma = 1$. \\ 

\subsection{Estimating the volatility}\label{subsec:volatility}

We assume next that the volatility of the underlying process is unknown, as it may occur in real situations. It is well known that, under model \eqref{eq:BB_SDE}, one can exactly compute the volatility if the price dynamics are continuously observed. However, investors in real life have to deal with discrete-time observations and thus they would have to estimate $\sigma$ to obtain the OSB. \\

We start by assuming that we have recorded the price values at the times $t_{0} = 0 < t_{1} < \cdots < t_{N - 1} < t_{N} = T$, for $N\in\mathbb{N}$, so at $t_{n}$, with $n\in\{0,1,\dots, N\}$, we have gathered a sample $(X_{t_i})_{i = 0}^{n}$ from the historical path of the Brownian bridge $(X_{t})_{t = 0}^{T}$ with $X_T = S$. 
From \eqref{eq:BB_explicit}, we have that
$$
X_{t_{i}}\mid X_{t_{i-1}} \sim \mathcal{N}\left(\mu(t_{i-1},X_{t_{i-1}},t_{i}), \nu_{\sigma}^{2}(t_{i-1},t_{i})\right),\ \ i = 1,\dots,n,
$$
and the log-likelihood function of the volatility takes the form	
\begin{align*}
	\ell(\sigma \mid (t_{i}, X_{t_i})_{i = 0}^{n}) = C - n\log(\sigma) - \frac{1}{2\sigma^{2}}\sum_{i=1}^{n}\left(\frac{X_{t_i} - \mu(t_{i-1},X_{t_{i-1}},t_{i})}{\nu_{1}(t_{i-1},t_{i})}\right)^{2},
\end{align*}
where $C$ is a constant independent of $\sigma$. The maximum likelihood estimator for $\sigma$ is given by
\begin{align*}
	\widehat{\sigma}_{n} &= \sqrt{\frac{1}{n}\sum_{i=1}^{n}\left(\frac{X_{t_i} - \mu(t_{i-1}, X_{t_{i-1}}, t_{i})}{\nu_{1}(t_{i-1}, t_{i})}\right)^{2}}.
\end{align*}
Under an equally spaced partition ($t_{i} = i\frac{T}{N}$, $i = 0, 1,\dots, N$), standard results on maximum likelihood (see \cite{dacunha1986estimation}) give that  
$$
\sqrt{n}(\widehat{\sigma}_{n}-\sigma)\rightsquigarrow\mathcal{N}\left(0,\frac{\sigma^{2}}{2}\right),
$$
when $n\rightarrow\infty$ (hence $N \rightarrow \infty$) and $T \rightarrow\infty$ such that $t_i - t_{i - 1} = T / N$ remains constant, with $i = 1,\dots, N$.

\subsection{Confidence intervals for the boundary}\label{subsec:confidence}

\begin{figure}[H]
	\vspace*{-0.5cm}
	\centering
	\includegraphics[scale = 1.1]{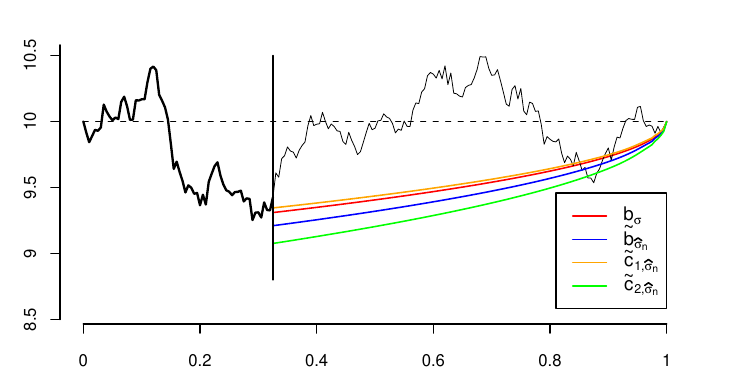}
	\caption{\small Inferring the boundary using one third ($n =  66$, $N = 200$) of the Brownian bridge path, for~$T = 1$, $S = 10$, $X_0 = 10$, $\lambda = 0$, and $\sigma = 1$. The solid curves represent the true boundary $b_\sigma$ (red curve), the estimated boundary $\tilde{b}_{\widehat{\sigma}_n}$ (blue curve), the upper confidence curve $\tilde{c}_{1, \widehat{\sigma}_n}$ (orange curve), and~the lower confidence curve $\tilde{c}_{2, \widehat{\sigma}_n}$ (green curve).
	}
	\label{fig:boundary_inference}
\end{figure}

\begin{figure}[h!]
	\vspace*{-0.5cm}
	\centering
	\begin{subfigure}[b]{0.47\textwidth}
		\includegraphics[width = \textwidth]{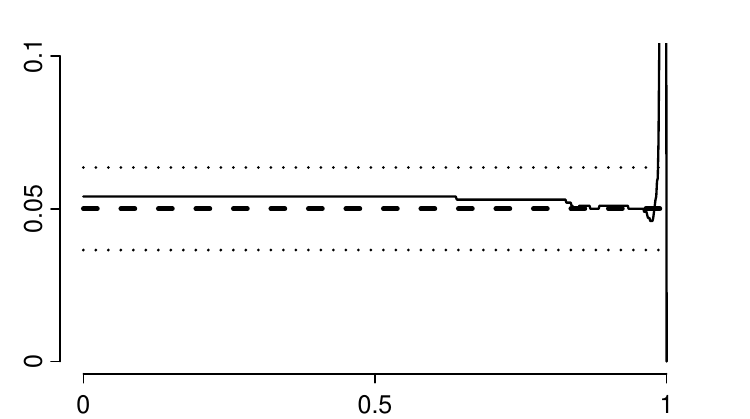}
		\subcaption[c]{One third of the paths $(n = 66)$.\label{fig.n.66}}
	\end{subfigure}
	\begin{subfigure}[b]{0.47\textwidth}
		\includegraphics[width = \textwidth]{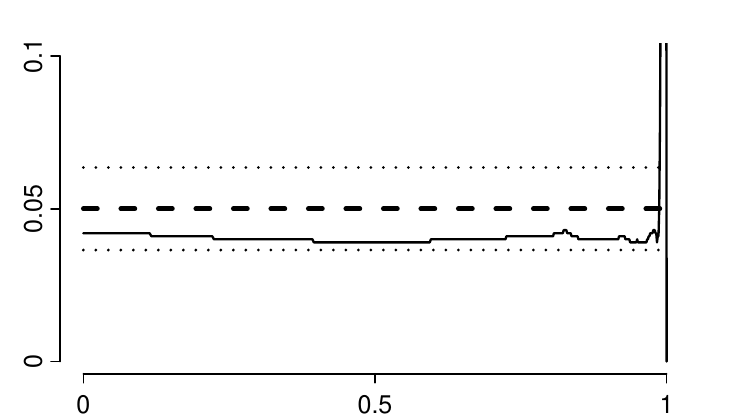}
		\subcaption[d]{Two thirds of the paths $(n = 133)$.\label{fig.n.133}}
	\end{subfigure}
	\vspace{6PT}
	
	\caption{%
		\small Pointwise proportion of trials, out of $M = 1000$, in which the true boundary does not belong to the interval delimited by the confidence curves. We use $S = 10$, $X_0 = 10$, $T = 1$, $\lambda = 0$,  $\sigma = 1$, and a significance level $\alpha = 0.05$, and a number $M = 1000$ of sample paths. For each path, one third (\textbf{a}) or two thirds (\textbf{b}) of the observations were used to compute $\sigma$ and then to estimate the confidence curves by~\eqref{eq:confidence_approx}. The continuous line represents the proportion of non-inclusions, the dashed line stands for $\alpha$, and the dotted lines are placed at the values $\alpha \pm z_{0.025}\sqrt{\frac{\alpha(1 - \alpha)}{M}}$. The spikes at $T = 1$ are numerical artifacts due to the null variance of $\tilde{b}_{\widehat{\sigma}_n}(T)$.}
	\label{fig:confidence_goodness}
\end{figure}

We present as follows the uncertainty propagated by the estimation of $\sigma$ to the computation of the OSB. 
In order to do so, we assume that the OSB is differentiable with respect to $\sigma$, so we are allowed to apply the delta method, under the previous asymptotic conditions. This entails that
\begin{align}\label{eq:b.Delta}
	\sqrt{n}(b_{\widehat{\sigma}_{n}}(t) - b_{\sigma}(t)) \rightsquigarrow \mathcal{N}\left(0, \left(\frac{\partial b_{\sigma}}{\partial\sigma}(\sigma,t)\right)^{2}\frac{\sigma^{2}}{2}\right),
\end{align} 
where $b_{\sigma}$ represents the OSB defined at \eqref{eq:b_volterra} associated with a process with volatility $\sigma$. Plugging-in the estimate $\widehat{\sigma}_n$ into \eqref{eq:b.Delta} gives the following asymptotic $100(1 - \alpha)\%$ (pointwise) confidence curves for $b_\sigma$:
\begin{align}\label{eq:confidence_approx}
	\left(c_{1,\widehat{\sigma}_{n}}(t),\ c_{2,\widehat{\sigma}_{n}}(t) \right) := \left(b_{\widehat{\sigma}_{n}}(t) \pm \frac{z_{\alpha/2}}{\widehat{\sigma}_{n}\sqrt{n/2}}\left|\frac{\partial b_{\sigma}}{\partial\sigma}(t)\Big\vert_{\sigma = \widehat{\sigma}_{n}}\right|\right),
\end{align}
where $z_{\alpha/2}$ represents the $\alpha/2$-upper quantile of a standard normal distribution. Algorithm \ref{alg:OSB} can be used to compute an approximation of the term $\frac{\partial b_{\sigma}}{\partial\sigma}(\cdot)$ by means of $(b_{\widehat{\sigma}_{n}+\varepsilon}(\cdot) - b_{\widehat{\sigma}_{n}}(\cdot))/\varepsilon$ for some small $\varepsilon>0$. We denote by $\left(\tilde{c}_{1,\widehat{\sigma}_{n}}(t),\ \tilde{c}_{2,\widehat{\sigma}_{n}}(t) \right)$ the approximation of the confidence interval \eqref{eq:confidence_approx} coming from this approach at $t\in[0, T]$. Through the paper, we use $\varepsilon = 10^{-2}$, as has been empirically checked to provide, along with $\delta=10^{-3}$ for Algorithm \ref{alg:OSB}, a good compromise between accuracy, stability, and computational speed in calculating the confidence curves. Figure \ref{fig:boundary_inference} illustrates, for one path of a Brownian bridge, how the boundary estimation and its confidence curves work. Figure \ref{fig:confidence_goodness} empirically validates the approximation of the confidence curves by marginally computing for each $t_n$ the proportion of trials, out of $M = 1000$, in which the true boundary does not belong to the interval delimited by the confidence curves.

The spikes visible in Figure \ref{fig:confidence_goodness} near the last point $t_{N} = T = 1$ indicate that the true boundary rarely lies within the confidence curves at those points. This  happens because the confidence curves have zero variance at the maturity date $T$ (actually $\tilde{c}_{1,\widehat{\sigma}_{n}}(T) =  \tilde{c}_{2,\widehat{\sigma}_{n}}(T) = S$), and the numerical approximation of $b(t_{N - 1})$ given at \eqref{eq:second-last_t_approx_1} is slightly biased. This affects the accuracy of the estimated boundary by frequently leaving the true boundary outside the confidence curves near maturity. This drawback is negligible in practice, since the estimated boundary and the confidence curves are very close to the true boundary in terms of absolute distance. 

\subsection{Simulations }\label{subsec:simul.study}

The ability to perform inference for the true OSB rises some natural questions: how much optimality is lost by $\tilde{b}_{\widehat{\sigma}_n}$ when compared to $b_\sigma$? How do the stopping strategies associated with the curves $\tilde{c}_{i, \widehat{\sigma}_n}$, $i = 1, 2$, compare with the one for $\tilde{b}_{\widehat{\sigma}_n}$? For example, a risk-averse (or risk-lover) strategy would be to consider the upper (lower) confidence curve $\tilde{c}_{1, \widehat{\sigma}_n}$ ($\tilde{c}_{2, \widehat{\sigma}_n}$) as the stopping rule, this being the most conservative (liberal) option within the uncertainty on estimating $b_\sigma$. A balanced strategy would be to consider the estimated boundary $\tilde{b}_{\widehat{\sigma}_n}$.\\

In the following, we investigate  how these stopping strategies behave, assuming $\sigma = 1$, $T = 1$, $S = 10$,  $X_0 = 10$, and $\lambda = 0$. We first estimate the payoff associated with each of them, and then we compare these payoffs with the one generated by considering the true boundary in its explicit form (see Remark \ref{rem:lambda0}). The choice of $\sigma = 1$ is not restrictive as it is enough to rescale time by $1/\sigma$ and space (i.e., the price values) by $1/\sqrt{\sigma}$. \\

\begin{figure}[H]
	\centering
	\vspace*{-1.5cm}
	\includegraphics[scale = 1.2]{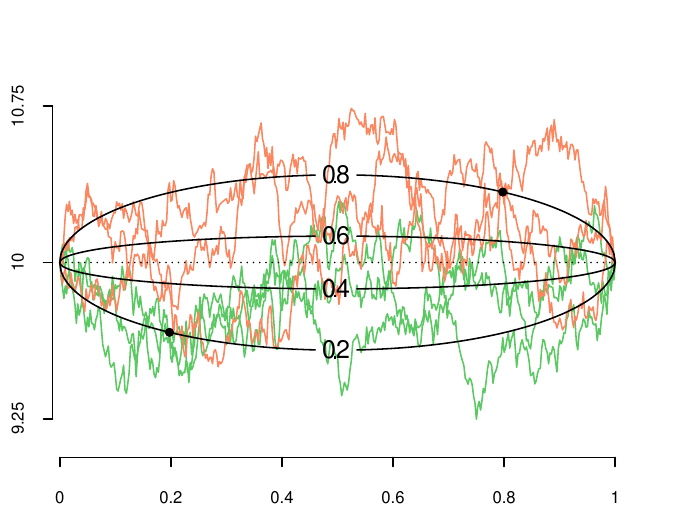}
	\caption{\small $X_{t}^{(q)}$ for $q= 0.2, 0.4, 0.6, 0.8$, where $X_{t}^{(q)}$ is the $q$-quantile of a $\mathcal{N}(0, t(1 - t))$, the marginal distribution at time $t$ of a Brownian bridge with unit volatility and $X_0 = X_1 = 10$. 
		The green and orange lines refer to the paths of $(X_t \mid X_{0.2} = X_{0.2}^{(0.2)})_{t = 0}^{1}$ and $(X_t \mid X_{0.8} = X_{0.8}^{(0.8)})_{t = 0}^{1}$ respectively, with~$X_{0.2}^{0.2} \approx 9.6649$ and $X_{0.8}^{0.8} \approx 10.3382$.}
	\label{fig:prc}
\end{figure}

To perform the comparison, we defined a subset of $[0, T]\times\mathbb{R}$ where the payoffs were computed. We~carried out the comparison along the pairs $\left(t_{i}, X_{t_{i}}^{(q)}\right)$, for $ i = 1,\dots, N$, $N = 200$, and $q = 0.2, 0.4, 0.6, 0.8$, where $t_{i} = i\frac{T}{N}$ and $X_{t_{i}}^{(q)}$ represents the $q$-quantile of the marginal distribution of the process at time $t_{i}$ that is $\mathcal{N}\left(0, \frac{t_i(T - t_i)}{T}\right)$ (see Figure \ref{fig:prc}).

For each $i$ and $q$, we generated $M = 1000$ different paths $(s_{j}, X_{s_j})_{j = 0}^{rN}$ of a Brownian bridge with volatility $\sigma = 1$ going from $(0, 0)$ to $(1, 0)$. Each path was sampled at times  $s_j = j\frac{T}{rn}$, for $j = 0, 1, \dots,rN$, for $r = 1$ and $r = 25$. The idea behind this setting is to tackle both the low-frequency scenario, which regards investors with access to daily prices or less frequent data, and the  high-frequency scenario, addressing high volumes of information as it happens to be when recording intraday prices. We forced each path to pass through $\left(t_{i}, X_{t_{i}}^{(q)}\right)$ (see Figure \ref{fig:prc}), and used the past $(s_{j}, X_{s_j})_{j=0}^{ri}$ of each trajectory to estimate the boundary and the confidence curves. The future $(s_{j}, X_{s_j})_{j=ri}^{rN}$ was employed to gather $M$ observations of the payoff associated with each stopping rule, whose means and variances are shown below in Figures \ref{fig:mean} and \ref{fig:var}, respectively.

\begin{figure}[H]
	\centering
	\begin{subfigure}[b]{0.45\textwidth}
		\includegraphics[width = \textwidth]{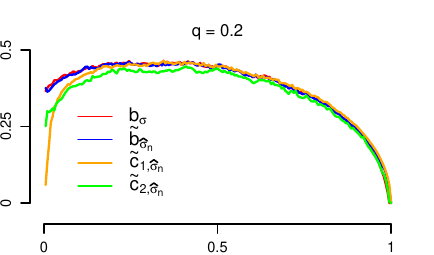}
	\end{subfigure}
	\begin{subfigure}[b]{0.45\textwidth}
		\includegraphics[width = \textwidth]{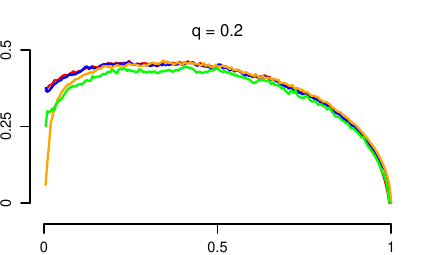}
	\end{subfigure}
	
	\begin{subfigure}[b]{0.45\textwidth}
		\includegraphics[width = \textwidth]{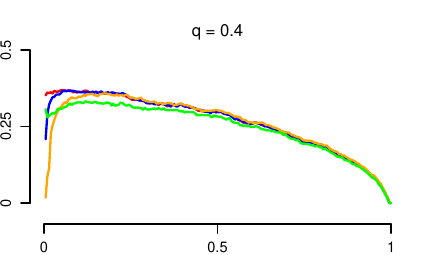}
	\end{subfigure}
	\begin{subfigure}[b]{0.45\textwidth}
		\includegraphics[width = \textwidth]{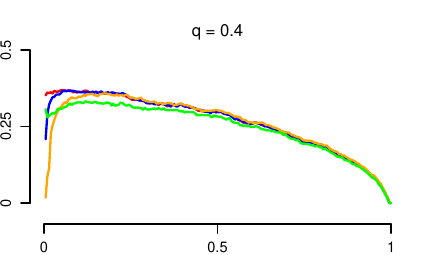}
	\end{subfigure}
	
	\begin{subfigure}[b]{0.45\textwidth}
		\includegraphics[width = \textwidth]{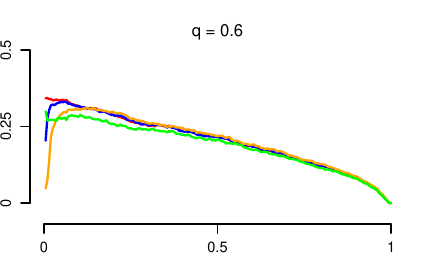}
	\end{subfigure}
	\begin{subfigure}[b]{0.45\textwidth}
		\includegraphics[width = \textwidth]{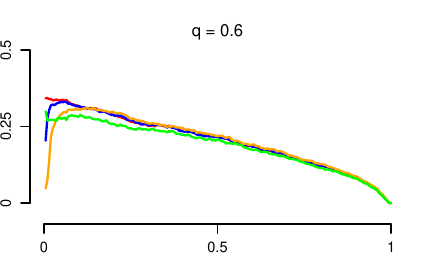}
	\end{subfigure}
	
	\begin{subfigure}[b]{0.45\textwidth}
		\includegraphics[width = \textwidth]{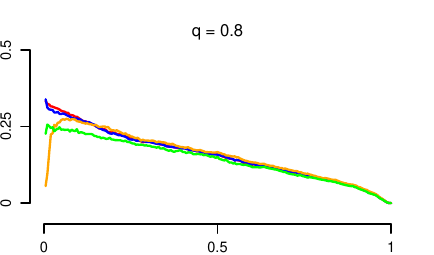}
	\end{subfigure}
	\begin{subfigure}[b]{0.45\textwidth}
		\includegraphics[width = \textwidth]{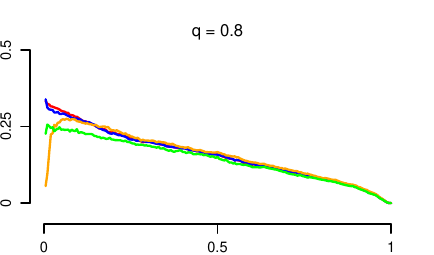}
	\end{subfigure}
	\caption{\small Mean of the payoff associated with: the true boundary $b_\sigma$ (red curve), the estimated boundary $\tilde{b}_{\widehat{\sigma}_n}$ (blue curve), the upper confidence curve $\tilde{c}_{1, \widehat{\sigma}_n}$ (orange curve), and the lower confidence curve $\tilde{c}_{2, \widehat{\sigma}_n}$ (green curve). The left column shows the low-frequency scenario ($r = 1$), while the right one stands for the high-frequency scenario ($r = 25$). We use $\sigma = 1$, $T = 1$, $S = 10$, $X_0 = 10$, and $\lambda = 0$.}
	\label{fig:mean}
\end{figure}

Figure \ref{fig:mean} shows the value functions associated with each stopping rule, the red curve being the one associated with the OSB. An important fact revealed by Figure \ref{fig:mean} is that in both the low and high-frequency scenarios the estimate $\tilde{b}_{\widehat{\sigma}_n}$ behaves almost indistinguishably to $b_{\sigma}$ in terms of the mean payoff after just a few initial observations.\\

Despite the variance payoff not being an optimized criterion in \eqref{eq:OSP_AmPut_BB}, it is worth knowing how it behaves for the three different stopping strategies, as it represents the risk associated with adopting each stopping rule as an exercise strategy. As expected, for any pair $(t, x)$, a higher stopping boundary implies a smaller payoff variance. 

\begin{figure}[h!]
	\centering
	\begin{subfigure}[b]{0.45\textwidth}
		\includegraphics[width = \textwidth]{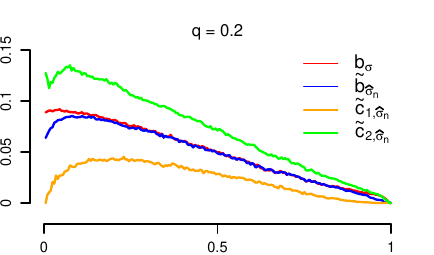}
	\end{subfigure}
	\begin{subfigure}[b]{0.45\textwidth}
		\includegraphics[width = \textwidth]{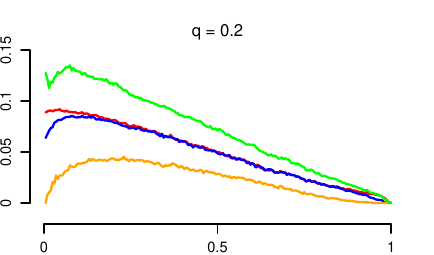}
	\end{subfigure}
	
	\begin{subfigure}[b]{0.45\textwidth}
		\includegraphics[width = \textwidth]{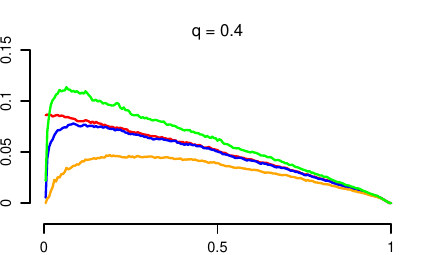}
	\end{subfigure}
	\begin{subfigure}[b]{0.45\textwidth}
		\includegraphics[width = \textwidth]{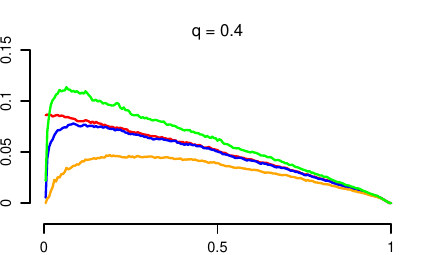}
	\end{subfigure}
	
	\begin{subfigure}[b]{0.45\textwidth}
		\includegraphics[width = \textwidth]{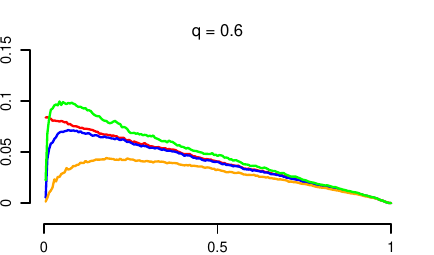}
	\end{subfigure}
	\begin{subfigure}[b]{0.45\textwidth}
		\includegraphics[width = \textwidth]{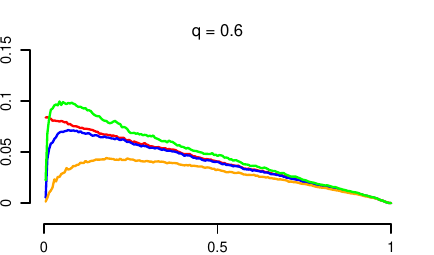}
	\end{subfigure}
	
	\begin{subfigure}[b]{0.45\textwidth}
		\includegraphics[width = \textwidth]{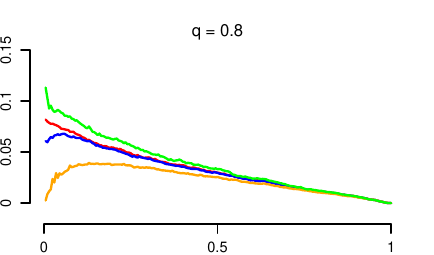}
	\end{subfigure}
	\begin{subfigure}[b]{0.45\textwidth}
		\includegraphics[width = \textwidth]{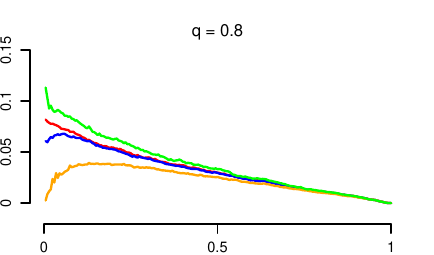}
	\end{subfigure}
	\caption{\small Variances of the payoff associated with: the true boundary $b_\sigma$ (red curve), the estimated boundary $\tilde{b}_{\widehat{\sigma}_n}$ (blue curve), the upper confidence curve $\tilde{c}_{1, \widehat{\sigma}_n}$ (orange curve), and lower confidence curve $\tilde{c}_{2, \widehat{\sigma}_n}$ (green curve). The left column shows the low-frequency scenario ($r = 1$), while the right one stands for the high-frequency scenario ($r = 25$). We use $\sigma = 1$, $T = 1$, $S = 10$, $X_0 = 10$, and $\lambda = 0$.}
	\label{fig:var}
\end{figure}

Figure \ref{fig:var} not only reflects this behavior by suggesting the upper confidence curve as the best stopping strategy, but also reveals that the variance does exhibit considerable differences for the stopping rules in the low-frequency scenario. These differences increase when the time gets closer to the initial point $t = 0$ and also when the quantile level $q$ decreases. In the high-frequency scenario, this effect is alleviated.\\

Figures \ref{fig:mean} and \ref{fig:var} also reveal that both the mean and the variance of the payoff associated with the estimated boundary $\tilde{b}_{\widehat{\sigma}}$ converge to the ones associated with the true boundary as more data are taken to estimate $\sigma$.\\

The pragmatic bottom line of the simulation study can be summarized in the following rules-of-thumb: if $15 < n < 1000$, it is advised to adopt the upper confidence curve as the stopping rule because $\tilde{c}_{1, \widehat{\sigma}_n}$ has almost the same mean payoff as all the other stopping rules while having considerable less variance; if $n \geq 1000$, the means and the variances of the payoff of the three stopping rules are quite similar, being the most efficient option to just assume $\tilde{b}_{\widehat{\sigma}_n}$ without computing the confidence curves. \\

For $n \leq 15$, the best candidate for the execution strategy is not obvious, and it would depend on which criterion is chosen to measure the mean-variance trade-off of the three strategies.

\section{Pinning at the strike and real data study}\label{sec:real_data}

In this section, we compare the performance of the optimal stopping strategy using the Brownian bridge model with a classical approach that uses the geometric Brownian motion \citep{peskir2005ontheamerican}. The latter does not take into account the pinning information of the asset’s price at maturity. We do so by a real data study analyzing various scenarios showing different degrees of intensity of pinning-at-the-strike.\\

The pinning behavior is more likely to take place among heavily traded options, as shown in~\cite{ni2005stock} and \cite{Nelken2001}. This is why we consider the options based on Apple and IBM expiring within the span of 11~January--18 September, in particular 8905 options for  Apple and 4833 for IBM. We denote by $M$ the total number of options of each company and, for the $j$-th option, we let $(X_{t_i}^{(j)})_{i = 0}^{N_j}$ be the 5-min tick close price of the underlying stock divided by the strike price $S_{j}$. In order to quantify the strength of the pinning effect, we define the \emph{pinning deviance} as $p_{j} := |X_{t_{N_j}}^{(j)} - 1|$, $ j = 1,\dots, M$. 
Therefore,~under~perfect pinning, we should expect $X_{t_{N_j}}^{(j)} = 1$ and~$p_j = 0$. \\

\begin{figure}[h!]
	\centering
	\begin{subfigure}[b]{0.45\textwidth}
		\includegraphics[width = \textwidth]{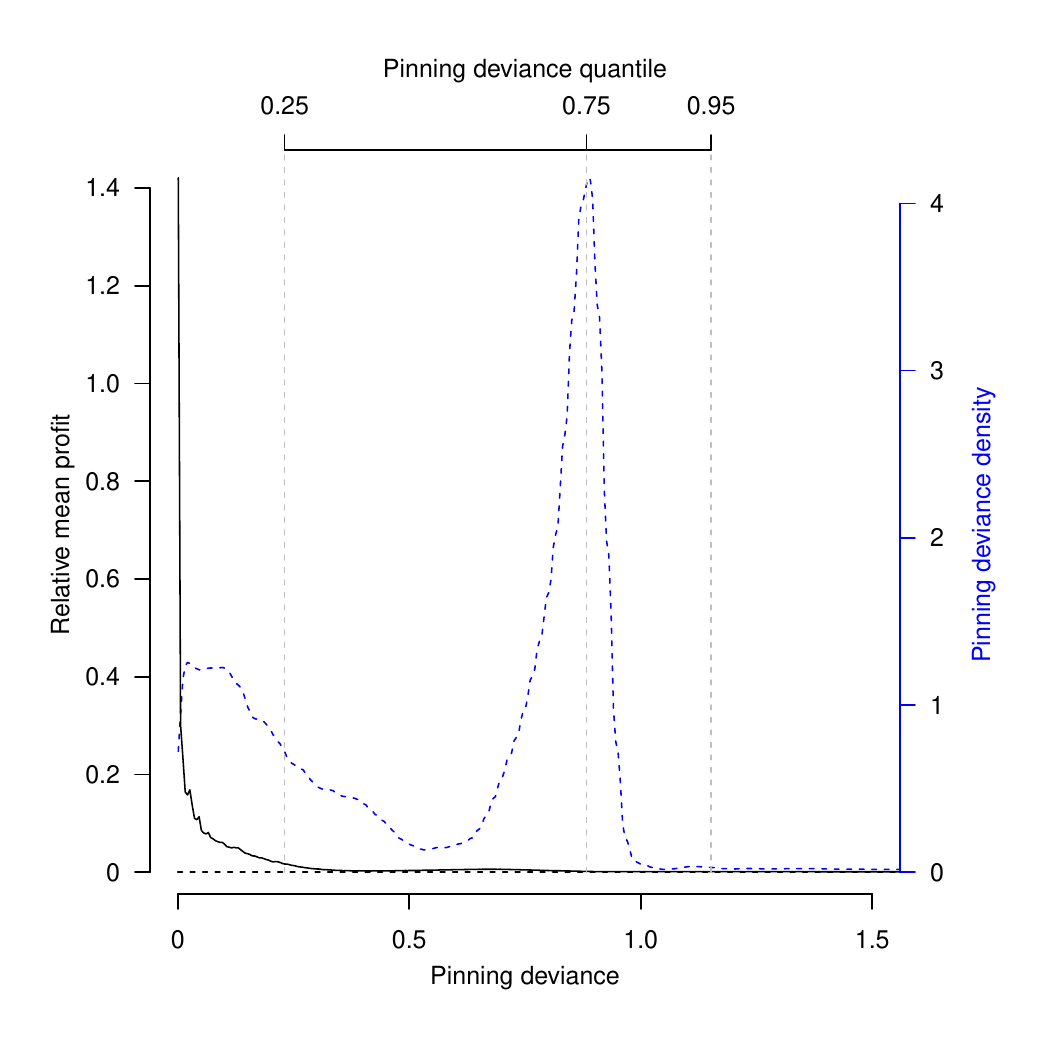}
		\caption{\small Apple.}
	\end{subfigure}
	\begin{subfigure}[b]{0.45\textwidth}
		\includegraphics[width = \textwidth]{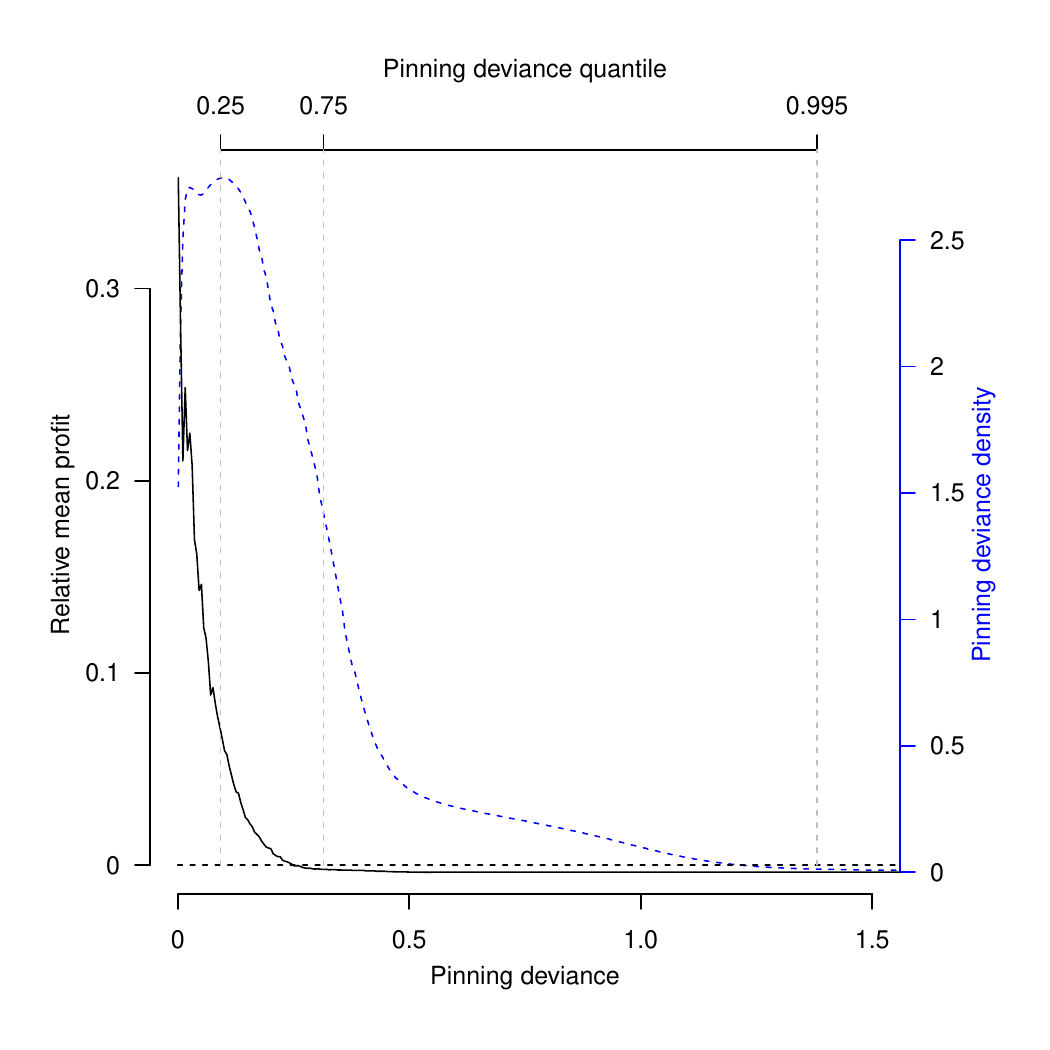}
		\caption{\small IBM.}
	\end{subfigure}
	\vspace{6PT}
	
	\caption{
		\small Results of the real data application. The black curve is the relative mean profit $\left(\mathrm{BB}(p) - \mathrm{GBM}(p)\right) / \mathrm{GBM}(p)$ for a pinning deviance $p$, while the blue dashed curve represents the kernel density estimation of the pinning deviances.
	}
	\label{fig:BB_vs_GBM}
\end{figure}

We perform the following steps in the real data application:

\begin{enumerate}
	\item We split each path $(X_{t_i}^{(j)})_{i = 0}^{N_j}$ into two subsets by using a factor $\rho \in \mathcal{P} = \{0.1, 0.2,\dots, 0.9\}$.
	We~call \emph{historical set} to the first $\rho 100 \%$ values of the prices $(X_{t_i}^{(j)})_{i = 0}^{\lfloor\rho N_j\rfloor}$
	and \emph{future set} to the remaining part (including the present value) $(X_{t_i}^{(j)})_{i = \lfloor\rho N_j\rfloor}^{N_j}$.
	Here, $j = 1,\dots, M$ while $1 - \rho$ represents the proportion of life time of the option.
	\item We use the historical set to estimate the volatility as described in Section \ref{subsec:volatility}.
	\item We compute the risk-free interest rate $\lambda_{j, \rho}$ as the $52$-week treasury bill rate (extracted from \cite{treasury}) held by the market when the split of $(X_{t_i}^{(j)})_{i = 0}^{N_j}$ was~done.
	\item We set the drift of the geometric Brownian motion to the risk-free interest rate such that the discounted process is a martingale.
	\item We compute the OSBs using Algorithm \ref{alg:OSB} for the Brownian bridge model \eqref{eq:b_volterra} and use the method exposed in \citet[page 12]{pedersen2002nonlinear} for the geometric Brownian motion model studied in \cite{peskir2005ontheamerican}. 
	Both numerical approaches are similar, the only subtle difference relies on the Brownian bridge requiring the last part of the integral to be computed as in Algorithm \ref{alg:OSB}, while the geometric Brownian motion needs no special treatment. The OSBs are computed with $S = 1$ (the stock prices were previously normalized by using the strike prices), $T = 1$ (all the maturity dates were standardized to $1$), and~$201$ nodes for the time partitions described in Section \ref{subsec:solving_the_fbe}.
	\item We compute the profit generated by optimally exercising the option within the remaining time by using the future set. This is done 
	by calculating $e^{-\lambda_{j, \rho} \tau_{\mathrm{BB}}^{j, \rho}}\Big(1 - X_{t_{\lfloor\rho N_j\rfloor} + \tau_{\mathrm{BB}}^{j, \rho}}^{(j)}\Big)$ and $e^{-\lambda_{j, \rho} \tau_{\mathrm{GBM}}^{j, \rho}}\Big(1 - X_{t_{\lfloor\rho N_j\rfloor} + \tau_{\mathrm{GBM}}^{j, \rho}}^{(j)}\Big)$, where $\tau_{\mathrm{BB}}^{j, \rho}$ and $\tau_{\mathrm{GBM}}^{j, \rho}$ are the OSTs associated, respectively, to the Brownian bridge and geometric Brownian motion strategies under the initial condition $\Big(t_{\lfloor\rho N_j\rfloor}, X_{t_{\lfloor\rho N_j\rfloor}}^{(j)}\Big)$.
	\item We compute the ``$\rho$-aggregated'' cumulative profit, as defined below, to measure the goodness of both models (BB stays for Brownian bridge while GBM stays for geometric Brownian motion):
	\begin{align*}
		\mathrm{BB}(p) &= \frac{1}{|\mathcal{P}||\mathcal{J}(p)|}\sum_{j \in \mathcal{J}(p)} \sum_{\rho \in \mathcal{P}} e^{-\lambda_{j, \rho} \tau_{\mathrm{BB}}^{j, \rho}}\Big(1 - X_{t_{\lfloor\rho N_j\rfloor} + \tau_{\mathrm{BB}}^{j, \rho}}^{(j)}\Big) , \\
		\mathrm{GBM}(p) &= \frac{1}{|\mathcal{P}||\mathcal{J}(p)|}\sum_{j \in \mathcal{J}(p)} \sum_{\rho \in \mathcal{P}} e^{-\lambda_{j, \rho} \tau_{\mathrm{GBM}}^{j, \rho}}\Big(1 - X_{t_{\lfloor\rho N_j\rfloor} + \tau_{\mathrm{GBM}}^{j, \rho}}^{(j)}\Big),
	\end{align*}
	where $\mathcal{J}(p):=\{j = 1,\dots, M : p_j < p\}$, and $|\mathcal{P}|$ and $|\mathcal{J}(p)|$ are the number of elements in $\mathcal{P}$ and $\mathcal{J}(p)$, respectively.
	\item We finally compute the relative mean profit $\left(\mathrm{BB}(p) - \mathrm{GBM}(p)\right) / \mathrm{GBM}(p)$.
	\item We plot the pinning deviances $p$ versus the relative mean profit (see Figure \ref{fig:BB_vs_GBM}).
\end{enumerate}

The Brownian bridge model behaves better than the geometric Brownian motion for options with low pinning deviance. This advantage fades away as we take distance from an ideal pinning-at-the-strike scenario, that is, when the pinning deviance increases. While the Brownian bridge model outperforms the geometric Brownian motion when applied to the Apple options along the whole dataset, when we consider the IBM options, the advantage is only present in $60\%$ of the options with lower pinning deviances.

\begin{Remark} 
	Besides using the OSB, we also considered in the analysis the confidence curves described in Section \ref{subsec:confidence}. However, since this is a high-frequency sampling scenario, both confidence curves provided almost indistinguishable results and were omitted to avoid redundancy. 
\end{Remark}

\begin{Remark}
	We did not consider the prices to buy the options when computing the profits in Figure \ref{fig:BB_vs_GBM}, as we are interested in when it is optimal to exercise the option rather than in whether it is profitable to buy the option held.
\end{Remark}

It is clear that an application of the Brownian bridge model is profitable in the presence of pinning-at-the-strike effect. However, it is far from being trivial to know beforehand if a stock will pin or not. Even if pinning forecasting is not the scope of this paper (for a systematic treatment, we refer to~\cite{avellaneda2003market}, \cite{jeannin2008modeling}, and \cite{avellaneda2012mathematical}), we provide some basic evidence about the possibility to predict the appearance of the pinning effect  by means of the trading volume of the options associated with a stock. For that, we study the association between the pinning deviances $(p_j)_{j = 1}^{M}$ and the number of open contracts for a given option that we call the \textit{Open Interest} (OI). In particular, we compute the \emph{weighted OI} for options expiring during the year 2017. Its definition is $\mathrm{wOI}_j := \sum_{k = 0}^{K_j}w_{j, k} o_{j, k}$, where $o_{j, k}$ is the OI of the $j$-th option at day $k$ after it was opened, $K_j$ is the total number of days the option remained available, and the weights $w_{j, k} := e^{-(1 - k/K_j)} / \sum_{i = 0}^{K_j} e^{-(1 - i/K_j)}$, $j = 1, \dots, M$, place more importance to OIs closer to the maturity date. We highlight that the $\mathrm{wOI}$ is an observable quantity. The Spearman's rank correlation coefficient between the wOI and the pinning deviances  scored $-0.5932$ for Apple and $-0.4281$ for IBM, thus~revealing a significant ($p$-values $< 10^{-16}$) positive dependence between $\mathrm{wOI}$ and the pinning~strength.

\section{Pinning at any point}\label{sec:pinning-at-any-point}

A recurring assumption in our work is that the pinning point coincides with the strike price of the option. We address next on what should be expected when relaxing this assumption.\\

A pinning point different from the strike prices would be desirable as it would increase the adaptability of the model to fit specific real situations. However, this extra flexibility makes the problem substantially more complex. For instance, the arguments used to state that the OSB is monotone in Proposition \ref{pr:b_existence} do not work anymore. Actually, empirical evidence suggests that this property does not hold anymore. Assuming that all regularity conditions still hold and that is possible to apply the extended version of the Itô's formula, then following the same arguments of Section \ref{sec:pricing} one would obtain the integral equation
\begin{align}\label{eq:b_volterra_any_pinning}
	b(t) = S - e^{-\lambda (T - t)}(S - A)^+ - \int_t^{T} K_{\sigma, \lambda}(t, b(t), u, b(u))\,\mathrm{d}u,
\end{align}
where $A$ and $S$ are the pinning point and the strike price respectively, and where 
\begin{equation*}
	\begin{aligned}
		K_{\sigma, \lambda}(t, x_1, u, x_2) &:= e^{-\lambda (u - t)}\left[\left(\frac{S}{T - u} + \lambda A - \left(\frac{1}{T - u} + \lambda\right)\mu(t, x_1, u)\right)\Phi(z_{\sigma}(t, x_1, u, x_2))\right. \\
		&\hspace{0.3cm}\left. +\ \nu_{\sigma}(t, u)\phi(z_{\sigma}(t, x_1, u, x_2))\right].
	\end{aligned}
\end{equation*}

\begin{figure}[h!]
	\vspace*{-0.5cm}
	\centering
	\includegraphics[scale = 1.1]{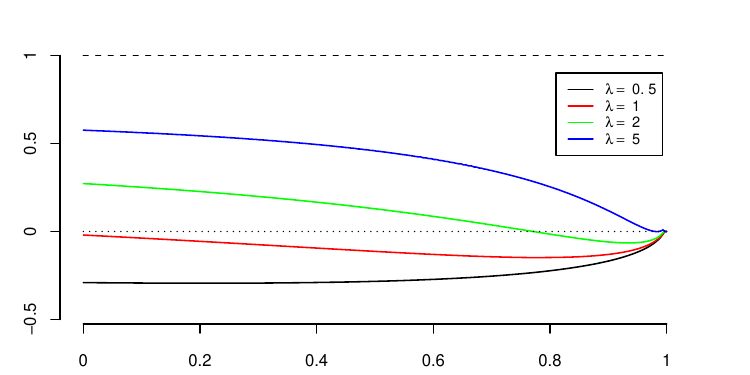}
	\caption{\small Numerical solution of the integral Equation \eqref{eq:b_volterra_any_pinning} for different values of $\lambda$. The dotted line represents the pinning point ($A = 0$) while the dashed line indicates the strike price ($S = 1$).}
	\label{fig:any_pinning}
\end{figure}

Figure \ref{fig:any_pinning} shows the optimal stopping boundary numerically obtained from \eqref{eq:b_volterra_any_pinning} for different discounting rates, suggesting that the OSB is not monotone.

Without monotonicity, it is much harder to prove other properties like the continuity of the boundary and the smooth fit condition. Some continuity results for the boundary for time homogeneous and non-homogeneous diffusions are proved in \cite{de_angelis_note_2013} and \cite{peskir_continuity_2019}, respectively, while \cite{cox_embedding_2015} addresses the smooth fit condition for a broad range of cases.\\

Some works go a step forward and try to solve a similar problem with a random pinning point, but in such cases different kinds of simplifications are required. For example, in \cite{ekstrom_optimal_2020}, the authors work in a non-discounted scenario with the identity as the gain function and gives bounds for the value function with a general pinning point distribution. Another good example is \cite{Leung2018}, where gain functions of class $C^2$ are considered.

\section{Conclusions}\label{sec:concluding_remarks} 

In this work we solved the problem of optimally exercising an American put option, in the presence of discount, by modeling the stock price with a Brownian bridge terminating at the strike price. The OSP was translated into a free-boundary problem shown to have a unique solution within a certain class of functions. We used a recursive fixed point algorithm to compute the OSB in practice, corroborating its accuracy in the case without discount. Using a maximum likelihood estimation for the volatility, we computed pointwise confident curves around the estimated OSB, and we analyzed some correlated alternative stopping rules. The simulation study done for the non-discounted case showed that the lower confidence curve is the most appealing stopping decision because of the resulting reduced variance of the optimal profit. Finally, we performed a real data study that empirically concluded that the Brownian bridge model behaves considerably better than a classical geometric Brownian motion when the stock prices exhibit the pinned-at-the-strike effect.\\

Our model not only requires an accurate insider information on the final value, but also is limited to the case in which the final value coincides with the strike price. A natural extension would be to find the optimal strategy for the case when the terminal point is different from the strike price. We~gave empirical evidence suggesting that in general the OSB is no longer monotone, resulting in a considerably more challenging problem. Another interesting extension would be to use a model that does not admit negative values for the stock price like, for instance, the exponential of a Brownian bridge or a geometric Brownian bridge.

\section*{Supplementary materials}\label{supplement}

The following supplementary materials are available online at \cite{AmOpBB_GitHub}. R scripts: \texttt{0-simulations.R}, \texttt{1-boundary\_computation\_BB-AmPut\_discount.R}, \texttt{2-inference\_BB.R}, \newline\texttt{3-simulation\_study.R}, and \texttt{4-figures.R}. RData files: \texttt{Prop1.RData}, \texttt{Prop2.RData}, \newline\texttt{payoff\_N200\_ratio1.RData}, and \texttt{payoff\_N200\_ratio25.RData}.

\section*{Acknowledgements}

The first author acknowledges support from projects MTM2017-85618-P and MTM2015-72907-EXP from the Spanish Ministry of Economy, Industry and Competitiveness, and the European Regional Development Fund.
The second author acknowledges support from projects PGC2018-097284-B-I00, IJCI-2017-32005, and MTM2016-76969-P  from the same funding agencies. The third author is supported by a scholarship from the Department of Statistics of Carlos III University of Madrid.

\appendix

\section{Main Proofs}\label{sec:sup_A}


\begin{proof}[Proof of Proposition \ref{pr:b_existence}]\label{proof:pr:b_existence}
	Take an admissible pair $(t, x)$ satisfying $x \geq S$ and $t < T$, and consider the stopping time $\tau_\varepsilon := \inf\left\{0\leq s\leq T - t : X_{t + s} \leq S - \varepsilon \mid X_t = x \right\}$ (assume for convenience that $\inf\{\emptyset\} = T - t$), for $\varepsilon > 0$. Notice that $\mathbb{P}_{t,  x}\left[\tau_\varepsilon < T - t\right] > 0$, which implies that $V(t, x) \geq \mathbb{E}_{t,x}\left[e^{-\lambda\tau_\varepsilon}G(X_{t + \tau_\varepsilon})\right] > 0 = G(x)$, from where it comes that $(t, x)\in C$.\\
	
	Define $b(t):= \sup\left\{x\in\mathbb{R} : (t, x) \in D\right\}$. The above arguments guarantee that $b(t) < S$ for all $t\in[0, T)$, and we get from (\ref{eq:OSP_AmPut_BB}) that $b(T) = S$. Furthermore, from \eqref{eq:OSP_AmPut_BB}, it can be easily noticed that, as $\lambda$ increases, $V(t, x)$ decreases. Therefore, $b(t)$ increases, and, since $b(t)$ is known to be finite for all $t$ when $\lambda = 0$ (see Remark \ref{rem:put-call}), then we can guarantee that $b(t) > -\infty$ for all values of $\lambda$.\\
	
	Notice that, since $D$ is a closed set, $b(t)\in D$ for all $t\in [0, T]$. In order to prove that $D$ has the form claimed in Proposition \ref{pr:b_existence}, let us take $x < b(t)$ and consider the OST $\tau^* = \tau^*(t, x)$. Then,~relying~on \eqref{eq:OSP_AmPut_BB}, \eqref{eq:BB_explicit}, and \eqref{eq:Val_fun}, we get
	\begin{align}\label{eq:VminusV_"spacewise"}
	V(t, x) - V(t, b(t)) &\leq \mathbb{E}_{t, x}\left[e^{-\lambda\tau^*}G(X_{t + \tau^*})\right] - \mathbb{E}_{t, b(t)}\left[e^{-\lambda\tau^*}G(X_{t + \tau^*})\right] \\
	&\leq \mathbb{E}_{t, 0}\left[\left(X_{t + \tau^*} + b(t)\frac{T - t - \tau^*}{T - t} - X_{t + \tau^*} - x\frac{T - t - \tau^*}{T - t}\right)^+\right] \nonumber \\
	&= (b(t) - x)\mathbb{E}\left[\frac{T - t - \tau^*}{T - t}\right] \nonumber \\
	&\leq b(t) - x, \nonumber
	\end{align}
	where we used the relation
	\begin{align}\label{eq:G_ineq}
	G(a) - G(b) \leq (b - a)^+,
	\end{align}
	for all $a, b \in \mathbb{R}$, for the second inequality. Since $V(t, b(t)) = S - b(t)$, we get from the above relation that $V(t, x) \leq S - x = G(x)$, which means that $(t, x) \in D$ and therefore $\left\{(t, x)\in[0,T]\times\mathbb{R}: x \leq b(t)\right\} \subset D$. On the other hand, if $(t, x) \in D$, then $x \geq b(t)$, which proves the reverse inclusion.\\
	
	Take now $t, t' \in [0, T]$ and $x \in \mathbb{R}$ such that $t' < t$ and $(t, x) \in C$. Then, since the function $t\mapsto V(t, x)$ is non-increasing for all $x \in \mathbb{R}$ (see \ref{pr:V.iv} from Proposition \ref{pr:V}), $V(t', x) \geq V(t, x) > G(x)$, i.e., $(t', x) \in C$. Hence, $b$ is non-decreasing.\\
	
	Finally, in order to prove the right-continuity of $b$, let us fix $t\in (0, T)$ and notice that, since $b$ is non-decreasing, then $b(t^+) \geq b(t)$. On the other hand, as $D$ is a closed set and $(t + h, b(t + h)) \in D$ for all $0 < h \leq T - t$, then $(t^+, b (t^+)) \in D$ or, equivalently, $b(t^+) \leq b(t)$.
\end{proof}


\begin{proof}[Proof of Proposition \ref{pr:V}]\label{proof:pr:V}
	
	\ref{pr:V.i} Half of the statement relies on the results obtained in \citet[Section 7.1]{goran-optimal} relative to the Dirichlet problem. It states that $\partial_tV + \mathbb{L}_{X}V = \lambda V$ on $C$. In addition,  it indicates how to prove that $V$ is $\mathcal{C}^{1, 2}$ from a solution of the parabolic Partial Differential Equation (PDE)
	\begin{align*}
	\partial_tf + \mathbb{L}_{X}f - \lambda f &= 0 && \text{in } R, \\
	f &= V && \text{on } \partial R,
	\end{align*} 
	where $R\in C$ is a sufficiently regular region. If we consider $R$ as an open rectangle, since $V$ is continuous by \ref{pr:V.v} below, and both $\mu$ and $\sigma$ are locally H\"older continuous, then the above PDE has a unique solution (see \citet[Theorem 9, Section 3]{Friedman1983}). Finally, since $V(t, x) = G (x) = S - x$ for all $(t, x)\in D$, $V$ is $\mathcal{C}^{1,2}$ also on $D$.\\
	
	\ref{pr:V.ii} We easily get the convexity of $x\mapsto V(t,x)$ by plugging-in \eqref{eq:BB_explicit} into \eqref{eq:OSP_AmPut_BB}. To prove \eqref{eq:V_x}, let us fix an arbitrary point $(t,x)\in[0,T]\times\mathbb{R}$, and consider $\tau^* = \tau^*(t,x)$ and $\varepsilon > 0$. Since $\tau_\varepsilon \rightarrow \tau$ a.s., by~arguing similarly to \eqref{eq:VminusV_"spacewise"}, we get
	\begin{align}\label{eq:V_x<=}
	\varepsilon^{-1}(V(t, x + \varepsilon) - V(t, x)) & \geq -\mathbb{E}\left[e^{-\lambda\tau^*}\frac{T - t - \tau^*}{T - t}\right],
	\end{align}
	where the limit is valid due to the dominated convergence theorem. For $\varepsilon < 0$, the reverse inequality emerges, giving us, after taking $\varepsilon \rightarrow 0$, the relation $\partial_x^-V(t, x) \leq -\mathbb{E}\left[e^{-\lambda\tau^*}\frac{T - t - \tau^{*}}{T-t}\right] \leq \partial_x^+V(t, x)$, which, due to the continuity of $x\mapsto \partial_xV(t, x)$ on $(-\infty, b(t))$ and on $(b(t), \infty)$ for all $t\in[0 , T]$ ($V$ is $\mathcal{C}^{1,2}$ on $C$ and on $D$), turns into $\partial_xV(t, x) = -\mathbb{E}\left[e^{-\lambda\tau^*}\frac{T-t-\tau^{*}}{T-t}\right]$ for all $(t, x)$ where $t\in[0, T]$ and $x\neq b(t)$. For~$x = b(t)$, Equation \eqref{eq:V_x} also holds and it turns into the smooth fit condition \ref{eq:smooth.fit} later proved.\\
	
	Furthermore, since $\mathbb{P}_{t,x}\left[\tau^{*}<T-t\right] > 0$ by Lemma \ref{lm:aux1}, \eqref{eq:V_x} shows that $\partial_xV < 0$ and therefore $x\mapsto V(t,x)$ is strictly decreasing for all $t\in[0,T]$.\\
	
	\ref{eq:smooth.fit} Take a pair $(t, x) \in [0, T)\times\mathbb{R}$ lying on the OSB, i.e., $x = b(t)$, and consider $\varepsilon > 0$. Since~$(t, x)\in D$ and $(t, x + \varepsilon)\in C$, we have that $V(t, x) = G(x)$ and $V(t, x + \varepsilon) > G(x + \varepsilon)$. Thus, taking into account the inequality \eqref{eq:G_ineq}, we get $\varepsilon^{-1}(V(t, x + \varepsilon) - V(t, x)) > \varepsilon^{-1}(G(x + \varepsilon) - G(x)) \geq -1$, which, after taking $\varepsilon\rightarrow 0$ turns into $\partial_x^+V(t, x)\geq -1$. On the other hand, consider the OST $\tau_{\varepsilon} := \tau^*(t, x + \varepsilon)$ and follow arguments similar to \eqref{eq:VminusV_"spacewise"} to get
	\begin{align}
	\varepsilon^{-1}(V(t, x + \varepsilon) - V(t, x)) & \leq -\mathbb{E}\left[e^{-\lambda\tau_\varepsilon}\frac{T - t - \tau_\varepsilon}{T - t}\right],
	\end{align}
	which, alongside the fact that $\tau_\varepsilon \rightarrow 0$ a.s. and using the dominated convergence theorem gives that $\partial_x^+V(t, x)\leq -1$. Therefore, $\partial_x^+V(t, b(t)) = -1$ for all $t\in[0,T)$. Since $V = G$ in $D$, it follows straightforwardly that $\partial_x^-V(t, b(t)) = -1$ and hence the smooth fit condition holds.\\
	
	\ref{pr:V.iv} Notice that, due to \ref{pr:V.i}, alongside \eqref{eq:inf_gen}, \eqref{eq:BB_explicit}, and \eqref{eq:V_x}, and, recalling that $x\mapsto V(t, x)$ is convex (and therefore $\partial_{x^2}V \geq 0$), we get that
	\begin{align*}
	\partial_tV(t, x) \leq \lambda V(t, x) - \frac{T - t - x}{T - t}\partial_xV(t, x) = -\mathbb{E}\left[e^{-\lambda\tau^*}(x - S)\frac{T - t - \tau^*}{T - t}\left(\lambda + \frac{x - S}{T - t}\right)\right].
	\end{align*}
	Therefore, $\partial_tV \leq 0$ on the set $C_S := [0, T)\times [S, \infty) \subset C$. \\
	
	For some small $\varepsilon > 0$, denote by $(X^{[t, T]}_s)_{s\geq0}^{T - t + \varepsilon}$ a process such that, for $s\in[0, T - t]$, it behaves as the Brownian bridge $X^{[t, T]}$, and the remaining part stays constant at the value $S$, i.e.,  $X^{[t, T]}_s = S$ for $s\in[T - t, T - t + \varepsilon]$. Let $\mu^{[t, T]}$ be its drift and define the process $(X^{[t-\varepsilon, T]}_s)_{s\geq0}^{T - t + \varepsilon} = X^{[t - \varepsilon, T]}$ with drift $\mu^{[t - \varepsilon, T]}$. Since $\mu^{[t, T]}(t, x) \geq \mu^{[t - \varepsilon, T]}$ whenever $x\leq S$, Theorem 1.1 from \cite{ikeda1977} guarantees that $X_s^{[t, T]} \geq X_s^{[t - \varepsilon, T]}\ \mathbb{P}_{t, x}$-a.s., for all $(t, x)$ and for all $s \leq \tau^S$, where $\tau^S := \inf\{s \in [0, T - t] : X_s^{[t, T]} > S \}$.\\
	
	Assume now that both processes start at $x\leq S$, and consider the stopping time $\tau^* = \tau^*(t, x)$. Therefore, since $V(t + s\wedge\tau^*, X^{[t, T]}_{s\wedge\tau^*})$ and $V(t + s, X^{[t-\varepsilon, T]}_s)$ are a martingale and a supermartingale (see~Section 2.2 from \cite{goran-optimal}), respectively, we have:
	\begin{align*}
	V(t, x) - V(t - \varepsilon, x) 
	\leq&\; \mathbb{E}\left[V(t + \tau^S\wedge\tau^*, X^{[t, T]}_{\tau^S\wedge\tau^*}) - V(t - \varepsilon + \tau^S\wedge\tau^*, X^{[t - \varepsilon, T]}_{\tau^S\wedge\tau^*})\right] \\
	\leq&\; \mathbb{E}\left[\mathbb{I}(\tau^*\leq\tau^S, \tau^* < T - t)e^{-\lambda\tau^*}(X^{[t-\varepsilon, T]}_{\tau^*} - X^{[t, T]}_{\tau^*})^+\right] \\
	& + \mathbb{E}\left[\mathbb{I}(\tau^*\wedge\tau^S = T - t)e^{-\lambda\tau^*}(X^{[t-\varepsilon, T]}_{T - t} - X^{[t, T]}_{T-t})^+\right] \\
	& + \mathbb{E}\left[\mathbb{I}(\tau^S\leq\tau^*, \tau^S < T - t)V(t + \tau^S, X^{[t, T]}_{\tau^S}) - V(t - \varepsilon + \tau^S, X^{[t - \varepsilon, T]}_{\tau^S})\right] \\
	\leq&\; \mathbb{E}\left[\mathbb{I}(\tau^S\leq\tau^*, \tau^S < T - t)V(t - \varepsilon + \tau^S, S) - V(t - \varepsilon + \tau^S, X^{[t - \varepsilon, T]}_{\tau^S})\right] 
	\leq 0.
	\end{align*}
	
	The second inequality comes after noticing that $\tau^*$ is not optimal for $X^{[t - \varepsilon, T]}$ and using \eqref{eq:G_ineq}; the~third inequality holds since $X^{[t-\varepsilon, T]}_{\tau^*} \leq X^{[t, T]}_{\tau^*}$ for $\tau^*\leq\tau^S$, $X^{[t-\varepsilon, T]}_{T - t} \leq X^{[t, T]}_{T-t}$ whenever $\tau^*\wedge\tau^S = T - t$, and the fact that $\partial_tV \leq 0$ on the set $C_S$; and the last inequality relies on the increasing behavior of $x\mapsto V(t, x)$. Finally, after dividing by $\varepsilon$ and taking $\varepsilon\rightarrow 0$, we get the claimed result. 
	This technique of comparing the processes starting at different times and same space value is based on a similar argument for a different gain function that has recently appeared in \cite{Tiziano}. \\
	
	\ref{pr:V.v} Let $(X_{t_i + s}^{[t_i, T]})_{s\geq 0}^{[0, T - t_i]}$ be a Brownian bridge going from $X_{t_i} = x$ to $X_T = S$ for any $x\in\mathbb{R}$, with~$i = 1, 2$. Notice that, according to \eqref{eq:BB_explicit}, the following holds:
	\begin{align}\label{eq:time_changed_BB}
	X_{t_2 + s'}^{[t_2, T]} \stackrel{d}{=} r^{1/2}X_{t_1 + s}^{[t_1, T]} + (1 - r^{1/2})(S - x)\frac{s}{T - t_1},
	\end{align}
	where $r = \frac{T - t_2}{T - t_1}$, $s \in [0, T - t_1]$, and $s' = sr \in [0, T - t_2]$.\\
	
	Take $0 \leq t_1 < t_2 < T$, consider $\tau_1 := \tau^*(t_{1}, x)$, and set $\tau_2 := \tau_1r$. Since $t\mapsto V(t,x)$ is decreasing for every $x\in\mathbb{R}$, then
	\begin{align*}
	0 &\leq V(t_1, x) - V(t_2, x) \\
	&\leq \mathbb{E}_{t_1, x}\left[e^{-\lambda\tau_1}G\left(X_{t_1 + \tau_1}^{[t_1, T]}\right)\right] - \mathbb{E}_{t_2, x}\left[e^{-\lambda\tau_2}G\left(X_{t_2 + \tau_2}^{[t_2, T]}\right)\right] \\
	& \leq \mathbb{E}\left[e^{-\lambda\tau_2}\left(G\left(X_{t_1 + \tau_1}^{[t_1, T]}\right) - G\left(X_{t_2 + \tau_2}^{[t_2, T]}\right)\right)\right] \\
	& \leq \mathbb{E}\left[\left(X_{t_2 + \tau_2}^{[t_2, T]} - X_{t_1 + \tau_1}^{[t_1, T]}\right)^+\right] \\
	& = \mathbb{E}\left[\left(\left(r^{1/2} - 1\right)\left(X_{t_1 + \tau_1}^{[t_1, T]} + (S - x)\frac{\tau_1}{T - t_1}\right)\right)^+\right] \\
	&\leq \left(\left(r^{1/2} - 1\right)\left(S + \mathbb{I}(x\leq S)(S - x)\right)\right)^+,
	\end{align*}
	where the first equality comes after applying \eqref{eq:time_changed_BB} and the last inequality takes place since $r < 1$ and $X_{t_1 + \tau_1}^{[t_1, T]}\leq S$.\\
	
	Hence, $V(t_1, x) - V(t_2, x) \rightarrow 0$ as $t_1 \rightarrow t_2$, i.e., $t \mapsto V(t, x)$ is continuous for every $x\in\mathbb{R}$. Thus, to address the continuity of $V$, it is sufficient to prove that, for a fixed $t$, $x\mapsto V(t, x)$ is uniformly continuous within a neighborhood of $t$. The latter comes after the following inequality, which comes right after applying similar arguments to those used in \eqref{eq:VminusV_"spacewise"}:
	\begin{align*}
	0 \leq V(t, x_1) - V(t, x_2) \leq (x_2 - x_1)\mathbb{E}\left[e^{-\lambda\tau^*}\frac{T - t - \tau^*}{T - t}\right] \leq x_2 - x_1,
	\end{align*}
	where $x_1, x_2 \in \mathbb{R}$ are such that $x_1 \leq x_2$ and $\tau^* = \tau^*(t, x_1)$.
\end{proof}


\begin{proof}[Proof of Proposition \ref{pr:b_continuity}]\label{proof:pr:b_continuity}
	We already proved the right-continuity of $b$ in Proposition \ref{pr:b_existence}, so this proof is devoted to prove its left-continuity.\\
	
	Let us assume that $b$ is not left-continuous. Therefore, as $b$ is non-decreasing, we can ensure the existence of a point $t_*\in (0, T)$ such that $b(t_*^-) < b(t_*)$, which allows us to take $x'$ in the interval $(b(t_*^-), b(t_*))$ and consider the right-open rectangle $\mathcal{R} = [t', t_*)\times[b(t_*^-), x']\subset C$ (see Figure \ref{fig:tik}), with~$t'\in(0, t_*)$.
	
	\begin{figure}[H]
		\centering
		\includegraphics[scale = 1.5]{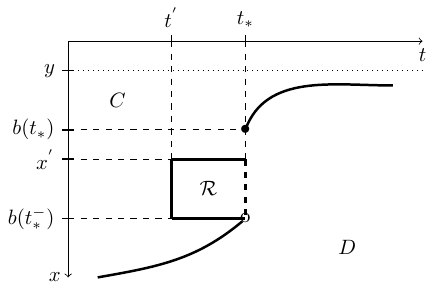}
		\caption{\small Graphical sketch of the proof of left continuity of $b$.}
		\label{fig:tik}
	\end{figure}
	
	Applying twice the fundamental theorem of calculus, using the fact that $(t, b(t))\in D$ for all $t\in[0,T]$, the~smooth fit condition \ref{eq:smooth.fit}, and the fact that $x\mapsto V(t, x)$ is $\mathcal{C}^{2}$ on $C$, we obtain
	\begin{align}\label{eq:b_continuity1}
	V(t, x) - G(x) = \int_{b(t)}^x\int_{b(t)}^u(\partial_{x^2}V(t, v) - \partial_{x^2}G(v))\,\mathrm{d}v\,\mathrm{d}u,
	\end{align}
	for all $(t,x)\in\mathcal{R}$. \\
	
	On the other hand, if we set $m := -\sup_{(t, x) \in \mathcal{R}}\partial_xV(t, x)$, then we readily obtain from \eqref{eq:V_x} that $m > 0$ (see Lemma \ref{lm:aux1}) which, combined with $\partial_tV + \mathbb{L}_XV = \lambda V$ on $C$ and $\partial_tV\leq 0$ on $C$ (\ref{pr:V.i} and \ref{pr:V.iv} in Proposition \ref{pr:V}), along with the fact that $V(t, x)\geq 0$ for all admissible pairs $(t, x)$, gives
	\begin{align}
	\partial_{x^2}V(t, x) = \frac{2}{\sigma^2}\left(\lambda V(t, x) - \frac{S - x}{T - t}\partial_xV(t, x) - \partial_tV(t, x)\right) \geq \frac{2m}{\sigma^2}\frac{S - x}{T-t} > 0 \label{eq:b_continuity2},
	\end{align}
	for all $(t,x) \in \mathcal{R}$. Therefore, by noticing that $\partial_{x^2}G(x) = 0$ for all $x\in(b(t_*^-), x')$ and plugging-in \eqref{eq:b_continuity2} into \eqref{eq:b_continuity1}, we get
	\begin{align*}
	V(t, x) - G(x) \geq \int_{b(t)}^x\int_{b(t)}^u \frac{2m}{\sigma^2}\frac{S - x}{T - t}\,\mathrm{d}v\,\mathrm{d}u 
	\geq \frac{2m}{\sigma^2}\frac{S - x}{T - t}\int_{b(t_*^-)}^x\int_{b(t_*^-)}^u\,\mathrm{d}v\,\mathrm{d}u 
	= \frac{2m}{\sigma^2}\frac{S - x}{T - t}(x - b(t_*^-))^2.
	\end{align*}
	
	Finally, after taking $t\rightarrow t_{*}$ on both sides of the above equation, we obtain $V(t_*, x) - G(x) > 0$ for all $x \in (b(t_{*}^{-}), b(t_*))$, which contradicts the fact that $(t_{*}, x)\in D$.
\end{proof}

\begin{proof}[Proof of Proposition \ref{pr:uniqueness}]\label{proof:pr:uniqueness}
	
	
	Assume we have a function $c:[0,T]\rightarrow \mathbb{R}$ that solves \eqref{eq:b_volterra} and define
	\begin{align}\label{eq:V^c}
	V^{c}(t, x) :=&\; \int_t^T e^{-\lambda (u - t)}\left(\frac{1}{T - u} + \lambda\right)\mathbb{E}_{t, x}\left[\left(S - X_u\right)\mathbbm{1}\left(X_u \leq c(u)\right)\right]\,\mathrm{d}u \\
	=&\; \int_t^T K_{\sigma, \lambda}(t, x, u, c(u))\,\mathrm{d}u, \nonumber
	\end{align}
	where $X = (X_s)_{s = 0}^{T}$ is a Brownian bridge with $\sigma$ volatility that ends at $X_T = S$, and $K_{\sigma, \lambda}$ is defined at \eqref{eq:K}. It turns out that $x\mapsto K_{\sigma, \lambda}(t, x, u, c(u))$ is twice continuously differentiable and therefore differentiating inside the integral symbol at \eqref{eq:V^c} yields $\partial_x V^c(t, x)$ and $\partial_{x^2} V^c(t, x)$, and furthermore ensures their continuity on $[0,T)\times\mathbb{R}$.\\
	
	
	Let us compute the operator $\partial_t + \mathbb{L}_X$ acting on the function $V^c$,
	\begin{align*}
	\partial_tV^c + \mathbb{L}_{X}V^{c}(t,x) &= \lim_{h\downarrow 0}\frac{\mathbb{E}_{t,x}[V^{c}(t + h, X_{t + h})] - V^{c}(t,x)}{h}.
	\end{align*}
	
	Define the function
	\begin{align}
	I(t, u, x_1, x_2) := e^{-\lambda (u - t)}\left(\frac{1}{T - u} + \lambda\right)\left(S - x_1\right)\mathbbm{1}\left(x_1 \leq x_2\right)
	\end{align}
	and notice that
	\begin{align*}
	\mathbb{E}_{t, x}[V^c(t + h, X_{t + h})] &= \mathbb{E}_{t, x}\left[\mathbb{E}_{t + h, X_{t + h}}\left[\int_{t + h}^{T}I(t + h, u, X_u, c(u))\,\mathrm{d}u\right]\right]\\
	&= \mathbb{E}_{t, x}\left[\mathbb{E}_{t, x}\left[\int_{t + h}^{T}I(t + h, u, X_u, c(u))\,\mathrm{d}u \ \Big\vert\ \mathcal{F}_{t + h}\right]\right]\\
	&= \mathbb{E}_{t, x}\left[\int_{t + h}^{T}I(t + h, u, X_u, c(u))\,\mathrm{d}u\right],
	\end{align*}
	where $(\mathcal{F}_s)_{s=0}^T$ is the natural filtration of $X$. Therefore,
	\begin{align*}
	\partial_tV^c + \mathbb{L}_{X}&V^{c}(t, x) \\
	&= \lim_{h\downarrow 0}\frac{\mathbb{E}_{t, x}\left[\int_{t + h}^{T}I(t + h, u, X_u, c(u))\,\mathrm{d}u\right] - \mathbb{E}_{t, x}\left[\int_{t}^{T}I(t, u, X_u, c(u))\,\mathrm{d}u\right]}{h}\\
	&=  \lim_{h\downarrow 0} \frac{1}{h}\mathbb{E}_{t, x}\left[\int_{t + h}^{T}\left(e^{\lambda h} - 1\right)I(t, u, X_u, c(u))\,\mathrm{d}u\right] - \lim_{h\downarrow 0} \frac{1}{h}\mathbb{E}_{t, x}\left[\int_{t}^{t + h}I(t, u, X_u, c(u))\,\mathrm{d}u\right]\\
	&= \lambda V(t, x) - (S - x)\left(\frac{1}{T - t} + \lambda\right)\mathbbm{1}(x \leq c(t)).
	\end{align*}
	
	From this result, alongside  with \eqref{eq:inf_gen} and the fact that $V^c$, $\partial_xV^{c}$, and $\partial_{x^2}V^{c}$ are continuous on $[0,T)\times\mathbb{R}$, we get the continuity of $\partial_t V^c$ on $C_1 \cup C_2$, where
	\begin{align*}
	C_{1} := \{(t,x) \in [0,T) \times \mathbb{R} : x > c(t)\},\ \ 
	C_{2} := \{(t,x) \in [0,T) \times \mathbb{R} : x < c(t)\}.
	\end{align*}
	
	Now, define the function $F^{(t)}(s, x) := e^{-\lambda s}V^{c}(t + s, x)$ with $s\in[0, T-t)$, $x\in\mathbb{R}$, and consider 
	\begin{align*}
	C_{1}^t := \{(s,x) \in C_1 : t \leq s < T \},\ \ 
	C_{2}^t := \{(s,x) \in C_2 : t \leq s < T \}.
	\end{align*}
	
	We claim that $F^{(t)}$ satisfies the \ref{lm:change_var2.iii-b} version of the hypothesis of Lemma \ref{lm:change_var2} taking $C = C_1^t$ and $D^{\circ} = C_2^t$. Indeed: $F^{(t)}$, $\partial_xF^{(t)}$, and $\partial_{x^2}F^{(t)}$ are continuous on $[0, T)\times\mathbb{R}$; it has been proved that $F^{(t)}$ is $\mathcal{C}^{1,2}$ on $C_1^t$ and $C_2^t$; we are assuming that $c$ is a continuous function of bounded variation; and $(\partial_tF^{(t)} + \mathbb{L}_X F^{(t)})(s, x) = -e^{-\lambda s}(S - x)\left(\frac{1}{T - t - s} + \lambda\right)\mathbbm{1}(x \leq c(t + s))$ is locally bounded on $C_1^t \cup C_2^t$.\\
	
	Thereby, we can use the \ref{lm:change_var2.iii-b} version of Lemma \ref{lm:change_var2} to obtain the following change-of-variable formula, which is missing the local time term because of the continuity of $F_x$ on $[0, T)\times\mathbb{R}$:
	\begin{align}
	e^{-\lambda s}&V^c(t + s,\ X_{t + s})\nonumber\\
	&= V^c(t, x) - \int_{t}^{t + s}e^{-\lambda(u - t)}(S - X_{u})\left(\frac{1}{T-u} + \lambda\right)\mathbbm{1}(X_u \leq c(u))\,\mathrm{d}u + M_{s}^{(1)}, \label{eq:V^c_changed}
	\end{align}
	with $M_{s}^{(1)} = \int_{t}^{t + s}e^{-\lambda (u - t)}\sigma \partial_xV^{c}(u, X_{u})\,\mathrm{d}B_{u}$. Notice that $(M_{s}^{(1)})_{s = 0}^{T - t}$ is a martingale under $\mathbb{P}_{t, x}$.\\
	
	
	In the same way, we can apply the \ref{lm:change_var2.iii-b} version of Lemma \ref{lm:change_var2} using the function $F(s, x) = e^{-\lambda s}G(X_{t + s})$, and taking $C = \{(s, x) \in [0, T - t)\times\mathbb{R} : x > S\}$ and $D^{\circ} = \{(s, x) \in [0, T - t)\times\mathbb{R} :$ $ x < S\}$, thereby getting
	\begin{align}\label{eq:G_changed}
	e^{-\lambda s}G(X_{t+s}) &= G(x) - \int_{t}^{t + s}e^{-\lambda(u - t)}(S - X_{u})\left(\frac{1}{T - u} + \lambda\right)\mathbbm{1}(X_u < S)\,\mathrm{d}u \\
	&\ \ \ - M_{s}^{(2)} + \frac{1}{2}\int_t^{t + s}e^{-\lambda (u - t)}\mathbbm{1}(X_u = S)\,\mathrm{d}l_s^S(X), \nonumber
	\end{align}
	where $M_{s}^{(2)} = \sigma\int_{t}^{t + s}e^{-\lambda (u - t)}\mathbbm{1}(X_u < S)\,\mathrm{d}B_{u}$, with $0 \leq s \leq T - t$, is a martingale under $\mathbb{P}_{t, x}$.\\
	
	
	Consider the following stopping time for $(t, x)$ such that $x \leq c(t)$:
	\begin{align}\label{eq:stopping_rho}
	\rho_c := \inf\left\{0 \leq s\leq T - t: X_{t + s} \geq c(t + s) \mid X_{t} = x\right\}.
	\end{align}
	
	In this way, along with assumption $c(t) < S$ for all $t \in (0, T)$, we can ensure that $\mathbbm{1}(X_{t + s} \leq c(t + s)) = \mathbbm{1}(X_{t + s} \leq S) = 1$ for all $s \in [0, \rho_c)$, as well as $\int_t^{t + s}e^{-\lambda (u - t)}\mathbbm{1}(X_u = S)\,\mathrm{d}l_s^S(X) = 0$.
	Recall that $V^{c}(t, c(t)) = G(c(t))$ for all $t\in[c,T)$ since $c$ solves \eqref{eq:b_volterra}. Moreover, $V^c(T, S) = 0 = G(S)$. Hence, $V^c(t + \rho_{c}, X_{t + \rho_c}) = G(X_{t + \rho_c})$. Therefore, we are able now to derive the following relation from Equations \eqref{eq:V^c_changed} and \eqref{eq:G_changed}:
	\begin{align*}
	V^c(t, x) &= \mathbb{E}_{t, x}[e^{-\lambda \rho_c}V^c(t + \rho_c,\ X_{t + \rho_c})] \\
	& \hspace{0.4cm} + \mathbb{E}_{t, x}\left[\int_{t}^{t + \rho_c}e^{-\lambda(u - t)}(S - X_{u})\left(\frac{1}{T-u} + \lambda\right)\mathbbm{1}(X_u \leq c(u))\,\mathrm{d}u\right] \\
	&= \mathbb{E}_{t, x}\left[e^{-\lambda\rho_c}G(X_{t + \rho_c})\right] \\
	&\hspace{0.4cm}  + \mathbb{E}_{t, x}\left[\int_{t}^{t + \rho_c}e^{-\lambda(u - t)}(S - X_{u})\left(\frac{1}{T-u} + \lambda\right)\mathbbm{1}(X_u \leq S)\,\mathrm{d}u\right] \\
	& = G(x).
	\end{align*}
	
	The vanishing of the martingales $M_{\rho_c}^{(1)}$ and $M_{\rho_c}^{(2)}$ comes after using the optional stopping theorem (see,~e.g.,~Section~3.2 from \cite{goran-optimal}). Therefore, we have just proved that $V^{c} = G$ on $C_{2}$.\\
	
	
	Now, define the stopping time
	\begin{align*}
	\tau_{c} := \inf\{0\leq u\leq T-t: X_{t + u} \leq c(t + u) \mid X_t = x\}
	\end{align*}
	and plug-in it into \eqref{eq:V^c_changed} to obtain the expression
	\begin{align*}
	V^c(t, x) &= e^{-\lambda \tau_c}V^c(t + \tau_c,\ X_{t + \tau_c}) \\
	&\ \ \ + \int_{t}^{t + \tau_c}e^{-\lambda(u - t)}(S - X_{u})\left(\frac{1}{T - u} + \lambda\right)\mathbbm{1}(X_u \leq c(u))\,\mathrm{d}u - M_{\tau_c}^{(1)}.
	\end{align*}
	
	Notice that, due to the definition of $\tau_{c}$, $\mathbbm{1}(X_{t+u} \leq c(t+u)) = 0$ for all $0\leq u < \tau_{c}$ whenever $\tau_c > 0$ (the case $\tau_c = 0$ is trivial). In addition, the optional sampling theorem ensures that $\mathbb{E}_{t, x}[M_{\tau_c}^{(1)}] = 0$. Therefore, the following formula comes after taking $\mathbb{P}_{t, x}$-expectation in the above equation and considering that $V^c = G$ on $C_2$:
	\begin{align*}
	V^{c}(t,x) = \mathbb{E}_{t, x}[e^{-\lambda\tau_c}V^{c}(t+\tau_{c}, X_{t + \tau_{c}})] = \mathbb{E}_{t,x}\left[e^{-\lambda\tau_c}G(X_{t+\tau_{c}})\right],
	\end{align*} 
	for all $(t,x)\in[0,T)\times\mathbb{R}$. Recalling the definition of $V$ from \eqref{eq:OSP_AmPut_BB}, the above equality leads to
	\begin{align}\label{eq:V^c<V}
	V^{c}(t,x)\leq V(t,x),
	\end{align}
	for all $(t,x)\in[0,T)\times\mathbb{R}$.\\
	
	
	Take $(t, x)\in C_{2}$ satisfying $x < \min\{b(t), c(t)\}$, where $b$ is the OSB for \eqref{eq:OSP_AmPut_BB}, and consider the stopping time $\rho_{c}$ defined as
	$$
	\rho_b := \inf\left\{0 \leq s\leq T - t: X_{t + s} \geq b(t + s) \mid X_{t} = x\right\}.
	$$
	
	Since $V = G$ on $D$, the following equality holds due to \eqref{eq:V_ito_formula} and from noticing that $\mathbbm{1}(X_{t + u} \leq b(t + u)) = 1$ for all $0\leq u < \rho_b$:
	\begin{align*}
	\mathbb{E}_{t, x}[e^{-\lambda\rho_b}V(t + \rho_b, X_{t + \rho_b})] = G(x) -  \mathbb{E}_{t, x}\left[\int_{t}^{t + \rho_b}e^{-\lambda(u - t)}(S - X_{u})\left(\frac{1}{T - u} + \lambda\right)\,\mathrm{d}u\right].
	\end{align*}
	
	On the other hand, we get the next equation after substituting $s$ for $\rho_b$ at \eqref{eq:V^c_changed}  and recalling that $V = G$ on $C_2$:
	\begin{align*}
	\mathbb{E}_{t, x}[e^{-\lambda\rho_b}V(t+ \rho_b, X_{t + \rho_b})] 
	= G(x) - \mathbb{E}_{t, x}\left[\int_{t}^{t + \rho_{c}}e^{-\lambda(u - t)}(S - X_{u})\left(\frac{1}{T - u} + \lambda\right)\mathbbm{1}(X_{u}\leq c(u))\,\mathrm{d}u\right].
	\end{align*}
	Therefore, we can use \eqref{eq:V^c<V} to merge the two previous equalities into
	\begin{align*}
	\mathbb{E}_{t,x}&\left[\int_{t}^{t + \rho_b}e^{-\lambda(u - t)}(S - X_{u})\left(\frac{1}{T - u} - \lambda\right)\mathbbm{1}(X_{u}\leq c(u))\,\mathrm{d}u\right]\\
	&\geq \mathbb{E}_{t, x}\left[\int_{t}^{t + \rho_b}e^{-\lambda(u - t)}(S - X_{u})\left(\frac{1}{T - u} - \lambda\right)\,\mathrm{d}u\right],
	\end{align*}
	meaning that $b(t) \leq c(t)$ for all $t\in[0,T]$ since $c$ is continuous. \\
	
	
	Suppose there exists a point $t\in(0, T)$ such that $b(t) < c(t)$ and fix $x\in(b(t), c(t))$. Consider the stopping time 
	\begin{align*}
	\tau_{b} := \inf\{0\leq u\leq T-t: X_{t + u} \leq b(t + u) \mid X_t = x\}
	\end{align*}
	and plug-in it both into \eqref{eq:V_ito_formula} and \eqref{eq:V^c_changed} replacing $s$ before taking the $\mathbb{P}_{t,x}$-expectation. We obtain
	\begin{align*}
	\mathbb{E}_{t, x}[e^{-\lambda \tau_b}V^c(t + \tau_b, X_{t + \tau_b})] &= \mathbb{E}_{t, x}[e^{-\lambda \tau_b}G(X_{t + \tau_b})] \\
	&= V^c(t, x) - \mathbb{E}_{t, x}\left[\int_{t}^{t + \tau_b}e^{-\lambda(u - t)}(S - X_{u})\left(\frac{1}{T - u} + \lambda\right)\mathbbm{1}(X_u \leq c(u))\,\mathrm{d}u\right]
	\end{align*}
	and
	\begin{align*}
	\mathbb{E}_{t, x}&[e^{-\lambda \tau_b}V(t + \tau_b, X_{t + \tau_b})] = \mathbb{E}_{t, x}[e^{-\lambda \tau_b}G(X_{t + \tau_b})] = V(t, x).
	\end{align*} 
	
	Thus, from \eqref{eq:V^c<V}, we get
	\begin{align*}
	\mathbb{E}_{t, x}\left[\int_{t}^{t + \tau_b}e^{-\lambda(u - t)}(S - X_{u})\left(\frac{1}{T - u} + \lambda\right)\mathbbm{1}(X_{u} \leq c(u))\,\mathrm{d}u\right] \leq 0.
	\end{align*}
	
	Using the fact that $x > b(t)$ and the time-continuity of the process $X$, we can state that $\tau_b > 0$. Therefore, the previous inequality can only happen if $\mathbbm{1}(X_{s} \leq c(s)) = 0$ for all $t\leq s\leq t + \tau_b$, meaning that $b(s) \geq c(s)$ for all $t\leq s\leq t + \tau_b$, which contradicts the assumption $b(t) < c(t)$.
\end{proof}

\begin{proof}[Proof of Theorem \ref{thm:OSP_sol}]\label{proof:thm:OSP_sol}
	Propositions \ref{pr:b_existence}--\ref{pr:b_continuity} give the required conditions to apply the Itô's formula extension exposed in the supplement to the function $F(s, x) = e^{-\lambda s}V(t + s, x)$, from where we get that
	\begin{align}\label{eq:V_ito_formula}
	e^{-\lambda s}V(t + s,X_{t + s}) &= V(t, X_t) + \int_0^s e^{-\lambda u}(\partial_tV + \mathbb{L}_{X}V - \lambda V)(t + u,X_{t + u})\,\mathrm{d}u \\
	&\ \ \ + \int_0^s\sigma e^{-\lambda u}\partial_xV(t + u,X_{t + u})\,\mathrm{d}B_{u}. \nonumber
	\end{align}
	
	Notice that the above formula is missing the local time term due to the continuity of $x\mapsto \partial_xV(t, x)$ for all $t\in [0, T]$.\\
	
	Recalling that $\partial_tV + \mathbb{L}_{X}V = \lambda V$ on $C$ and $(\partial_tV + \mathbb{L}_{X}V - \lambda V)(t, x) = -(S - x)\left(\frac{1}{T - t} + \lambda\right)$ for all $(t, x) \in D$, taking $\mathbb{P}_{t, x}$-expectation (causing the vanishing of the martingale term), setting $s = T - t$, and~making a simple change of variable in the integral, we get from \eqref{eq:V_ito_formula} the following pricing formula for the American put option:
	\begin{align}\label{eq:V_int}
	V(t, x) &= \int_t^Te^{-\lambda (u - t)}\left(\frac{1}{T - u} + \lambda\right)\mathbb{E}_{t, x}\left[\left(S - X_u\right)\mathbbm{1}\left(X_u \leq b(u)\right)\right]\,\mathrm{d}u.
	\end{align}
	
	We know from \eqref{eq:BB_explicit} that, for $u\in[t, T]$, $X_u^{[t, T]}\sim \mathcal{N}\left(\mu(t, x, u), \nu^2_{\sigma}(t, u)\right)$ under $\mathbb{P}_{t, x}$, where $\mu$ and $\nu_{\sigma}$ are given in \eqref{eq:BB_mean} and \eqref{eq:BB_var}, respectively. \\
	
	For any random variable $Y$, we have that $\mathbb{E}\left[Y\mathbbm{1}(Y \leq a)\right] = \mathbb{P}[Y \leq a]\mathbb{E}\left[Y\mid Y \leq a\right]$. In addition, if~$Y \sim \mathcal{N}(\mu, \nu^2)$, then $\mathbb{E}\left[Y \mid Y \leq a\right] = \mu - \frac{\nu\phi(z)}{\Phi(z)}$, where $z = \frac{a - \mu}{\nu}$, and $\phi$ and $\Phi$ denote, respectively, the~density and distribution functions. Then, the more tractable representation \eqref{eq:V_volterra} for $V$ follows. \\
	
	Since $V(t, x) = S - x$ for all $(t, x)\in D$, we can take $x \uparrow b(t)$ on both sides in \eqref{eq:V_volterra} in order to obtain the type two Volterra nonlinear integral Equation \eqref{eq:b_volterra} for the OSB $b$.\\
	
	Finally, due to Proposition \ref{pr:uniqueness}, we obtain that the solution of Equation \eqref{eq:b_volterra} is unique up to the regularity conditions considered in Theorem \ref{thm:OSP_sol}.
\end{proof}


\section{Auxiliary lemmas}\label{sec:sup_B}

\begin{Lemma}\label{lm:aux1}
	Let $(X_{t + s})_{s=0}^{T-t}$ be a Brownian bridge from $X_t$ to $X_T = S$ with volatility $\sigma$, where $t\in[0, T)$. Let $b$ be the OSB associated with the OSP
	\begin{align*}
	V(t,x)=\sup_{0\leq\tau\leq T-t}\mathbb{E}_{t,x}\left[e^{-\lambda\tau}G(X_{t+\tau})\right],
	\end{align*}
	with $G(x) = (G - x)^+$, and $\lambda \geq 0$. Then, $\sup_{(t,x)\in\mathcal{R}}\partial_xV(t,x) < 0$, where $\mathcal{R}$ is the set defined in the proof of Proposition \ref{pr:b_continuity}.
\end{Lemma}

\begin{proof}
	Take $0 < \varepsilon < 1$, let $\tau^{*} = \tau^{*}(t, x)$, and define 
	$$
	p(t,x):= \mathbb{P}\left[\tau^{*}\leq (T-t)(1-\varepsilon)\right].
	$$
	Notice that 
	\begin{align*}
	p(t, x)&= \mathbb{P}_{t, x}\left[\min_{0\leq s \leq (T - t)(1 - \varepsilon)}\{X_{t + s} - b(s)\} < 0\right] \\
	&\geq \mathbb{P}_{t, x}\left[\min_{0\leq s \leq (T - t)(1 - \varepsilon)}X_{t + s} < b(t)\right] \\
	&= \mathbb{P}\left[\min_{0\leq s \leq (T - t)(1 - \varepsilon)}\left\{(S - x)\frac{s}{T - t} + \sigma\sqrt{\frac{T - t - s}{T - t}}W_s\right\} < b(t) - x\right] \\
	&\geq \mathbb{P}\left[\min_{0\leq s \leq (T - t)(1 - \varepsilon)}\left\{\sqrt{\frac{T - t - s}{T - t}}W_s\right\} < \sigma^{-1}(b(t) - \max\{x, S\})\right] \\
	&= \mathbb{P}\left[\min_{0\leq s \leq (T - t)(1 - \varepsilon)}\left\{W_s\right\} < \varepsilon^{-1/2}\sigma^{-1}(b(t) - \max\{x, S\})\right] \\
	&= 2\mathbb{P}\left[W_{(T - t)(1 - \varepsilon)} < \varepsilon^{-1/2}\sigma^{-1}(b(t) - \max\{x, S\}\right],
	\end{align*}
	where the first inequality is justified since $b$ is non-decreasing (see Proposition \ref{pr:b_existence}), while the last equality comes after applying the \textit{reflection principle}. Therefore,
	\begin{align*}
	M := \inf_{(t,x)\in \mathcal{R}} p(t,x) & > 0.
	\end{align*}
	Finally, by using \eqref{eq:V_x}, we obtain the following relation for all $(t,x)\in\mathcal{R}$:
	\begin{align*}
	\partial_xV(t, x)  &\leq -e^{-\lambda(T - t)}\mathbb{E}\left[\frac{T - t - \tau^*}{T - t} \mathbbm{1}\left(\tau^{*}\leq (T-t)(1-\varepsilon)\right)\right]\  \\
	&\leq -e^{-\lambda(T - t)}\varepsilon p(t, x) \\
	&\leq -e^{-\lambda(T - t)}\varepsilon M < 0.
	\end{align*} 
\end{proof}

For the sake of completeness, we formulate the following change-of-variable result by taking Theorem 3.1 from \cite{peskir2005a} and changing some of its hypotheses according to Remark 3.2 from \cite{peskir2005a}. Specifically, the $\ref{lm:change_var2.iii-a}$ version of Lemma \ref{lm:change_var2} comes after changing, in \cite{peskir2005a}, (3.27) and (3.28) for the joint action of (3.26), (3.35), and (3.36). The $\ref{lm:change_var2.iii-b}$ version relaxes condition (3.35) into (3.37) in \cite{peskir2005a}. 

\begin{Lemma}\label{lm:change_var2}
	Let $X = (X_t)_{t=0}^{T}$ be a diffusion process solving the SDE
	\begin{align*}
	\,\mathrm{d}X_{t} = \mu(t, X_{t})\,\mathrm{d}t + \sigma(t, X_{t})\,\mathrm{d}B_{t},\ \ \ \ 0 \leq t \leq T,
	\end{align*} 
	in Itô's sense. Let $b:[0, T]\rightarrow\mathbb{R}$ be a continuous function of bounded variation, and let $F:[0, T] \times \mathbb{R} \rightarrow \mathbb{R}$ be a continuous function satisfying
	\begin{align*}
	F &\text{ is } \mathcal{C}^{1,2} \text{ on } C, \\
	F &\text{ is } \mathcal{C}^{1,2} \text{ on } D^{\circ},
	\end{align*}
	where $C = \{(t,x)\in [0,T]\times\mathbb{R} : x > b(t)\}$ and $D^{\circ} = \{(t,x)\in [0,T]\times\mathbb{R} : x < b(t)\}$.\\
	
	Assume there exists $t\in[0,T]$ such that the following conditions are satisfied:
	\begin{enumerate}[label=(\roman{*}), ref=(\textit{\roman{*}}),leftmargin=2.3em,labelsep=4mm]
		\item $\partial_tF + \mu \partial_xF + (\sigma^2 / 2)\partial_{x^2}F$ is locally bounded on $C \cup D^{\circ}$;
		\item the functions $s\mapsto \partial_xF(s, b(s)^\pm):= \partial_xF(s, \displaystyle{\lim_{h\to 0+} b(s)\pm h})$ are continuous on $[0, t]$;
		\item and either
		\begin{enumerate}[label=(iii-\alph{*}), ref=(\textit{iii-\alph{*}})]
			\item $x\mapsto F(s, x)$ is convex on $[b(s) - \delta, b(s)]$ and convex on $[b(s), b(s) + \delta]$ for each $s\in[0,t]$, with some $\delta > 0$, or, \label{lm:change_var2.iii-a}
			\item $\partial_{x^2}F = G_1 + G_2$ on $C\cup D^{\circ}$, where $G_1$ is non-negative (or non-positive) and $G_2$ is continuous on $\bar{C}$ and $\bar{D^{\circ}}$. \label{lm:change_var2.iii-b}
		\end{enumerate}
	\end{enumerate}
	
	Then, the following change-of-variable formula holds:
	\begin{align*}
	F(t, X_t) =&\; F(0, X_0) + \int_{0}^{t}(\partial_tF + \mu \partial_xF + (\sigma^2/2)\partial_{x^2}F)(s, X_s)\mathbbm{1}(X_s \neq b(s))\,\mathrm{d}s\ \\
	&+ \int_{0}^{t}(\sigma \partial_xF)(s, X_s)\mathbbm{1}(X_s \neq b(s))\,\mathrm{d}B_s \\
	&+ \frac{1}{2}\int_0^t(\partial_xF(s, X_s^+) - \partial_xF(s, X_s^-))\mathbbm{1}(X_s = b(s))\,\mathrm{d}l_s^b(X),
	\end{align*}
	where $\mathrm{d}l_s^b(X)$ is the local time of $X$ at the curve $b$ up to time $t$, i.e.,
	\begin{align}
	l_s^b(X) = \lim_{\varepsilon\rightarrow 0}\int_0^t\mathbbm{1}(b(s) - \varepsilon \leq X_s \leq b(s) + \varepsilon)\,\mathrm{d}\langle X, X\rangle_s,
	\end{align}
	where $\langle X, X\rangle$ is the predictable quadratic variation of $X$, and the limit above is meant in probability.
\end{Lemma}


\end{document}